\documentclass[a4paper,11pt]{article}
\usepackage[mathlines]{lineno}
\usepackage{jcappub}

\usepackage[table,svgnames,dvipsnames]{xcolor}

\usepackage[normalem]{ulem}
\usepackage{aas_macros}
\usepackage{subfigure}
\usepackage{graphicx}% Include figure files
\usepackage{graphics}
\usepackage{dcolumn}% Align table columns on decimal point
\usepackage{bm}% bold math
\usepackage{orcidlink}
\usepackage{cases}% for dealing with mathematics,
\usepackage{booktabs}
\usepackage{comment}
\usepackage{multirow}
\usepackage{makecell}
\usepackage{siunitx}
\usepackage{tabularx}
\usepackage{xspace}
\usepackage{soul} % can comment out at the end
\graphicspath{{figs/}}

\usepackage{orcidlink} % for author list
\usepackage{hanging} % for affiliations
\usepackage{arydshln} % for horizontal dashed line, \hdashline
\usepackage{makecell} % for splitting cell in table into multiple rows

\renewcommand{\arraystretch}{1.4}
\def\be{\begin{equation}}
\def\ee{\end{equation}}

\def\ba#1\ea{\begin{align*}#1\end{align*}}

\renewcommand{\emph}[1]{\textit{#1}}
\definecolor{RoyalBlue}{rgb}{0.25,.41,.88}
\definecolor{WildStrawberry}{HTML}{EE2967}
\definecolor{RedWine}{rgb}{0.743,0,0}
\definecolor{bittersweet}{rgb}{1.0, 0.44, 0.37}
\definecolor{burntorange}{rgb}{0.8, 0.33, 0.0}
\definecolor{midnightgreen}{rgb}{0.0, 0.29, 0.33}
\definecolor{otherblue}{rgb}{0.20, 0.73, 0.92}

\usepackage[nameinlink,noabbrev]{cleveref}
\crefname{equation}{Eq.}{Eqs.}
\crefname{section}{Section}{Sections}
\crefname{figure}{Figure}{Figures}
\crefname{table}{Table}{Tables}
\crefname{appendix}{Appendix}{Appendices}
\Crefname{figure}{Figure}{Figures}
\Crefname{equation}{Equation}{Equations}
\Crefname{section}{Section}{Sections}
\Crefname{table}{Table}{Tables}

\newcommand{\mksym}[1]{\ifmmode {\rm #1}\else #1\fi}
\newcommand{\dataplus}{\allowbreak+}

\newcommand{\leftparbox}[2]{\parbox{#1}{\begin{flushleft} #2 \end{flushleft}}}
\newcommand{\oneonesig}[4][5cm]{
\begin{equation}
\left.
#2 \quad\mbox{\text{\leftparbox{#1}{(#3)#4}}}
  \right.
\end{equation}
}
\newcommand{\onetwosig}[4][5cm]{
\begin{equation}
\left.
  #2 \quad\mbox{\text{\leftparbox{#1}{(95\,\%,~#3)#4}}}
  \right.
\end{equation}
}
\newcommand{\twoonesig}[4][\pbwidth]{
\begin{equation}
\left.
 \begin{aligned}
#2 \\ #3
 \end{aligned}
\ \right\} \ \ \mbox{\text{\leftparbox{#1}{#4}}}
\end{equation}
}

\newcommand{\threeonesig}[5][\pbwidth]{
\begin{equation}
\left.
 \begin{aligned}
#2 \\ #3 \\ #4
 \end{aligned}
\ \right\} \ \ \mbox{\text{\leftparbox{#1}{#5}}}
\end{equation}
}

\newcommand{\Om}{\Omega_\mathrm{m}}
\newcommand{\Ocdm}{\Omega_\mathrm{cdm}}
\newcommand{\Ob}{\Omega_\mathrm{b}}
\newcommand{\Ok}{\Omega_\mathrm{K}}

\newcommand{\Ode}{\Omega_\mathrm{de}}
\newcommand{\ns}{n_\mathrm{s}}
\newcommand{\nsten}{n_\mathrm{s10}}
\newcommand{\Neff}{N_{\mathrm{eff}}}

\newcommand{\lcdm}{$\Lambda$CDM} 
\newcommand{\wcdm}{$w$CDM} 
\newcommand{\wowacdm}{$w_0w_a$CDM} 
\newcommand{\lya}{Ly$\alpha$\xspace}

\newcommand{\sumnu}{\sum m_\nu}

\newcommand{\Planck}{\emph{Planck}}

\newcommand{\Seight}{S_8}

\newcommand{\hinvmpc}{\,h^{-1}{\rm Mpc}}
\newcommand{\hmpcinv}{\,h\,{\rm Mpc^{-1}}}
\newcommand{\kmsMpc}{\,{\rm km\,s^{-1}\,Mpc^{-1}}}
\newcommand{\eV}{{\,\rm eV}}

\newcommand{\FM}{Full Modeling}

\newcommand{\planckact}{\Planck+ACT}

\title{DESI 2024 VII:  Cosmological Constraints from the Full-Shape Modeling of Clustering Measurements}

\author{{DESI Collaboration}:}
\emailAdd{spokespersons@desi.lbl.gov}
\affiliation{Affiliations are in Appendix \ref{sec:affiliations}}

\author[1]{{A.~G.~Adame},}
\author[2]{{J.~Aguilar},}
\author[3]{{S.~Ahlen}\orcidlink{0000-0001-6098-7247},}
\author[4]{{S.~Alam}\orcidlink{0000-0002-3757-6359},}
\author[5,6]{{D.~M.~Alexander}\orcidlink{0000-0002-5896-6313},}
\author[7,8]{{C.~Allende~Prieto}\orcidlink{0000-0002-0084-572X},}
\author[2]{{M.~Alvarez},}
\author[9]{{O.~Alves},}
\author[2]{{A.~Anand}\orcidlink{0000-0003-2923-1585},}
\author[10,9]{{U.~Andrade}\orcidlink{0000-0002-4118-8236},}
% ...

\date{\today}

\abstract{
We present cosmological results from the measurement of clustering of galaxy, quasar and Lyman-$\alpha$ forest tracers from the first year of observations with the Dark Energy Spectroscopic Instrument (DESI Data Release 1). We adopt the full-shape (FS) modeling of the power spectrum, including the effects of redshift-space distortions, in an analysis which has been thoroughly validated in a series of supporting papers as summarised in \cite{DESI2024.V.KP5}. We combine the full-shape information with DESI's DR1 constraints from the baryon acoustic oscillations (BAO) of these tracers. In the flat \lcdm\ cosmological model, DESI (FS+BAO), combined with a baryon density prior from Big Bang Nucleosynthesis and a weak prior on the scalar spectral index, determines matter density to $\Om=0.2962\pm 0.0095$, and the amplitude of mass fluctuations to $\sigma_8=0.842\pm 0.034$. The addition of the cosmic microwave background (CMB) data tightens these constraints to $\Om=0.3056\pm 0.0049$ and $\sigma_8=0.8121\pm 0.0053$, while further addition of the the joint clustering and lensing analysis from the Dark Energy Survey Year-3 (DESY3) data further improves these measurements, and leads to a 0.4\% determination of the Hubble constant, $H_0 = (68.40\pm 0.27)\,\kmsMpc$. In models with a time-varying dark energy equation of state parametrised by $w_0$ and $w_a$, combinations of DESI (FS+BAO) with CMB and type Ia supernovae continue to show the preference, previously found in the DESI DR1 BAO analysis,  for $w_0>-1$ and $w_a<0$ with similar levels of significance. DESI data, in combination with the CMB, improve the upper limits on the sum of the neutrino masses relative to the case when only the DR1 BAO was available, giving $\sumnu < 0.071\eV$ at 95\% confidence. We finally constrain deviations from general relativity represented by two modified gravity parameters. DESI (FS+BAO) data alone measure the parameter that controls the clustering of massive particles, $\mu_0=0.11^{+0.45}_{-0.54}$, in agreement with the zero value predicted by general relativity.  The combination of DESI with the CMB and the clustering and lensing analysis from DESY3 constrains both modified-gravity parameters, giving $\mu_0 = 0.04\pm 0.22$ and $\Sigma_0 = 0.044\pm 0.047$, again in agreement with general relativity.  
}

\begin{document}
\maketitle
\flushbottom

%%%%%%%%%%%%%%%%%%%%%%%%%%%%%%%%%%%%%%%%%%%%%%%%%%%%%%%%%%%%
\section{Introduction}
\label{sec:intro}
%%%%%%%%%%%%%%%%%%%%%%%%%%%%%%%%%%%%%%%%%%%%%%%%%%%%%%%%%%%%

The large-scale structure (LSS) of the Universe, as probed by galaxy surveys and the intergalactic medium, has firmly established itself as a reliable probe of cosmology and fundamental physics. The three-dimensional clustering of tracers of the LSS --- galaxies, quasars, and \lya\ absorption signatures in quasar spectra ---  can be directly related to cosmological theory. This, in turn, can be used to constrain some of the most familiar quantities in cosmology, including the amount of dark matter and dark energy, the amplitude and spectral index of primordial density perturbations, spatial curvature, and neutrino mass. The progress in such clustering measurements over the last half century has been nothing short of remarkable \cite{1973ApJ...185..413P,1974ApJS...28...19P,Davis:1982gc,1988ApJ...332...44D,1993MNRAS.265..145B,20052dFBAO,2006PhRvD..74l3507T,Percival:2006gt,Reid:2009xm,Parkinson:2012vd}, and has resulted in percent-level constraints on some of the aforementioned cosmological parameters.

One prominent feature in the galaxy clustering correlation is the baryon acoustic oscillations (BAO),  an oscillatory signature which appears as  ``wiggles" in the galaxy power spectrum, or a single localised peak in the galaxy correlation function.  The scale of the BAO feature is determined by the sound horizon at the baryon drag epoch, and its observation via tracers at a given redshift $z$ contains information about the ratio of this scale to distance measures ($D_A(z)$ and $c/H(z)$ in the directions perpendicular and parallel to the line-of-sight, respectively), thus containing key cosmological information. More information, however, is available in the ``full-shape'' of the clustering signal, specifically the measured  power spectrum $P(k, z)$ over a range of wavenumbers $k$ and tracer redshifts $z$ or, equivalently, the correlation function $\xi(r, z)$ where $r$ is the comoving separation. 

Notably, the dependence of the full-shape clustering signal on redshift $z$ informs us about the growth of cosmic structure (e.g.\ \cite{Peebles:1980yev,Liddle:2000cg,Huterer:2022dds}), which in turn is very sensitive to the properties of dark energy and modified gravity, and to the total matter content of the universe. The growth-rate constraints allow data to test the underlying theory of gravity at cosmological scales and discriminate between models that share the same expansion history, see e.g.  \cite{Koyama:2015vza,Joyce:2016vqv,Ishak:2018his,2021JCAP...11..050A}. Because DESI measurements are sensitive to both the geometrical quantities and the growth of density perturbations, they are particularly well-suited to supply tests of dark energy and modified gravity. 
The full-shape clustering signal also contains information about the amplitude and shape of the primordial power spectrum, and hence provide information complementary to that from the cosmic microwave background (CMB) measurements. 

The above-mentioned long history of measurements of galaxy clustering has, over the past decade or so, been reinvigorated with the data from the Baryon Oscillation Spectroscopic Survey (BOSS) \cite{2013AJ....145...10D} which has been part of the third phase of the Sloan Digital Sky Survey (SDSS-III; \cite{2011AJ....142...72E}), and its extension eBOSS \cite{2016AJ....151...44D}. Full-shape analyses of galaxy and quasar clustering in BOSS have been carried out by the BOSS \cite{BOSS:2016wmc,2017MNRAS.466.2242B,2017MNRAS.467.2085G,2017MNRAS.464.1640S,2017MNRAS.469.1369S} and eBOSS \cite{eBOSS:2020yzd,2021MNRAS.500..736B,2020MNRAS.498.2492G,2020MNRAS.499.5527T,2021MNRAS.501.5616D,2021MNRAS.500.1201H,2020MNRAS.499..210N} collaborations, as well as independent teams who typically studied BOSS and/or eBOSS data \cite{Ivanov:2019pdj,DAmico:2019fhj,Troster:2019ean,Chen:2021wdi,Kobayashi:2021oud,Philcox:2021kcw,Brieden:2022lsd,Noriega:2024lzo,Schoneberg:2022ggi,Chudaykin:2022nru,Donald-McCann:2023kpx,McDonough:2023qcu,Gsponer:2023wpm}. The tools developed in these analyses have enabled reliable extraction of cosmological information from clustering. 

The Dark Energy Spectroscopic Instrument (DESI) is  the first Stage-IV galaxy survey in operation \cite{Snowmass2013.Levi,DESI2016a.Science,DESI2023a.KP1.SV,DESI2023b.KP1.EDR}. It is conducting a spectroscopic five-year survey over $14{,}200$ square degrees that will collect spectra of about 40 million galaxies and quasars \cite{DESI2016b.Instr,DESI2022.KP1.Instr,FocalPlane.Silber.2023,Corrector.Miller.2023,SurveyOps.Schlafly.2023,Spectro.Pipeline.Guy.2023}. DESI targets five separate tracers: low-redshift galaxies from the Bright Galaxy Survey (BGS) \cite{BGS.TS.Hahn.2023}, luminous red galaxies (LRG) \cite{LRG.TS.Zhou.2023}, emission line galaxies (ELG) \cite{ELG.TS.Raichoor.2023}, quasars (QSO) \cite{QSO.TS.Chaussidon.2023}, and the \lya\ forest features in quasar spectra \cite{TS.Pipeline.Myers.2023}. DESI's deep redshift coverage, $0<z<4$, will enable it to map out the expansion history and the growth of cosmic structure to high precision.  The principal scientific goals of DESI are to obtain tight constraints on dark energy, neutrino mass, and primordial non-Gaussianity. This is complemented by a tremendous amount of other science that is being carried out using data from the DESI instrument.

This paper is part of a series discussing key cosmology results from the first year of observations from DESI, which is based on DESI Data Release 1 (DR1; \cite{DESI2024.I.DR1}). This is the second paper that focuses on key cosmological parameter measurements from DESI DR1; in the first paper \cite{DESI2024.VI.KP7A}, we presented cosmological measurements from the information in baryon acoustic oscillations in DESI DR1 data, based on the analysis of galaxy and quasar clustering \cite{DESI2024.III.KP4}, and that in the \lya\ forest data \cite{DESI2024.IV.KP6}. In this paper, we significantly extend those results by complementing the BAO information with the ``full-shape" analysis which models the overall clustering signal of DESI tracers across time and space, and report the resulting cosmological constraints from the combined BAO + full-shape analysis. 

The data behind the analysis, and the plans for their release, are presented in \cite{DESI2024.I.DR1}, while the galaxy/quasar samples are discussed in detail in \cite{DESI2024.II.KP3}.  The large-scale structure catalogs are fully described in \cite{DESI2024.II.KP3,KP3s15-Ross}. The DESI DR1 galaxy full-shape analysis, its detailed pipeline choices, the study of systematics, and the cosmological constraints on the $\Lambda$CDM model from DESI DR1 galaxy full-shape alone and its combination with BAO are all presented in \cite{DESI2024.V.KP5}. A further detailed analysis of modified gravity models is presented in \cite{KP7s1-MG}. Moreover, a number of technical details, as well as in-depth discussions and justification arguments for our analysis choices are provided in a series of supporting papers.  Specifically, \cite{KP5s2-Maus,KP5s3-Noriega,KP5s4-Lai,KP5s5-Ramirez,KP5s7-Findlay} provide details and validation of the perturbation theory codes that we use to analyse the (pre-reconstruction) galaxy power spectrum, while \cite{KP5s1-Maus,DESI2024.V.KP5} show the level of agreement between the codes in a series of controlled settings with simulated and synthetic data vectors.  These papers also discuss the role of priors and the projection effects that can arise when presenting high-dimensional posteriors marginalised to show constraints in lower-dimensional parameter spaces of interest.  The covariance matrices are described and validated in \cite{KP4s6-Forero-Sanchez,KP4s7-Rashkovetskyi,KP4s8-Alves}.  Our systematic error budget relies on studies which are presented in \cite{KP5s6-Zhao,KP5s7-Findlay,KP5s8-Findlay} and summarised in \cite{DESI2024.V.KP5}.  Throughout the analysis we have made use of a series of mock catalogs described in detail in \cite{KP3s8-Zhao}. Note that the constraints on primordial NG will be presented separately in \cite{ChaussidonY1fnl}.

%%%%%%%%%%%%%%%%%%%%%%%%%%%%%%%%%%%%%%%%%%%%%%%%%%%%%%%%
\section{Data and methodology}
\label{sec:data}
%%%%%%%%%%%%%%%%%%%%%%%%%%%%%%%%%%%%%%%%%%%%%%%%%%%%%%%%

In this Section we describe the essential inputs to the cosmological analysis --- data and methodology. In \cref{sec:FS_meas} we describe the data, full-shape measurement methodology, and the blinding procedure that we applied to the measurements. In \cref{sec:ext}, we describe the external data that we optionally combine with DESI in the analysis. Finally, \cref{sec:modeling} describes theoretical modeling, as well as the likelihood analysis and other details of our cosmological inference pipeline. 

\subsection{DESI Full-Shape measurements}
\label{sec:FS_meas}

\subsubsection{DESI DR1 data}
\label{sec:DESI_data}

The DESI data that we use are described in \cite{DESI2024.II.KP3}.  They are derived from the redshifts and positions of over 4.7 million unique galaxies and QSOs over a $\sim$$7{,}500$ square degree footprint\footnote{Note that the sky coverage for individual tracers may be substantially lower than $\sim$$7{,}500$ sq.~deg.\ due to masks and cuts; see \cite{DESI2024.II.KP3} for details.} covering the redshift range $0.1 < z < 2.1$.  These discrete tracers are broken into four target classes:  300,017 galaxies from the magnitude-limited bright galaxy survey (BGS); 2,138,600 luminous red galaxies (LRG); 1,415,707 emission line galaxies (ELG)\footnote{The DESI DR1 sample contains a total of 2,432,022 ELGs in two redshift bins, but the ELGs in the lower redshift bin (1,016,365 objects) did not pass the systematics checks \cite{DESI2024.V.KP5}, so we do not use them in the cosmological analysis.} and 856,652 quasars (QSO) (see Table 1 of \cite{DESI2024.V.KP5}). 
These tracers are split into six redshift bins: one bin with the BGS ($0.1<z<0.4$), three bins with the LRGs ($0.4<z<0.6$, $0.6<z<0.8$, and $0.8<z<1.1$), one bin with the ELGs ($1.1<z<1.6$), and one redshift bin with the QSOs ($0.8<z<2.1$).
These objects are assembled into large-scale structure catalogs, and the power spectrum in each redshift bin is subsequently computed as discussed below; see \cite{DESI2024.II.KP3} and references therein for all details.

In addition to the discrete tracers described above, DESI also uses the spectra of distant QSOs to measure large-scale structure in the intergalactic medium (i.e., the \lya\ forest).  Measurements of the 3D correlation function of the DR1 \lya\ forest data are presented in \cite{DESI2024.IV.KP6}.
At present we only use the baryon acoustic oscillation information in the large-scale clustering of the \lya\ forest to constrain the background geometry \cite{DESI2024.IV.KP6}, and do not provide a measurement of growth.  For this reason the \lya\ forest measurements only enter  via their contribution to constraining the expansion history.

\subsubsection{DESI full-shape measurements}

The goal of our analysis is to extract cosmological information beyond the BAO feature from the measurements \cite{DESI2024.II.KP3,DESI2024.V.KP5} of the full-shape clustering of DESI tracers. To that effect, we measure the first few multipole moments of the Fourier-space tracer power spectra relative to the line-of-sight to the observer -- the monopole, quadrupole and hexadecapole -- which quantify the information imprinted by redshift-space distortions (although we limit our analysis here to the first two of these multipoles). These measurements are obtained with the estimator from \cite{yamamoto2006}. The power spectrum measurements are obtained from the galaxy catalogs (``data") and from synthetically-generated catalogs with random distribution of points (``randoms") to which we assign the same selection as for the data, including assigning weights (to points) that account for systematic corrections, and those that implement the Feldman-Kaiser-Peacock (FKP) optimal weighting scheme \cite{FKP}. We also use the random catalog to compute the window matrix \cite{Beutler:2021eqq,KP3s5-Pinon} that relates the measured power spectrum multipoles to the theory power spectrum prediction.

Data and random catalogs are constructed as described in \cite{DESI2024.II.KP3}; they are both masked for the presence of bright objects, lack of or bad imaging data and spectroscopic observations, and target priorities. Fiber assignment results in variations of the observed density of tracers; this effect is corrected by applying the completeness weights at the catalog level. Despite this correction, fiber assignment impacts the two-point statistics at small angular separations, which we consequently remove from the power spectrum estimation~\cite{KP3s5-Pinon}. Both the small-angle structure in the masks and the small-scale angular cuts result in a window matrix that has contributions extending to very small scales; we then ``rotate" \cite{KP3s5-Pinon} our power spectrum measurement, covariance matrix, and window matrix to make the latter more diagonal. Imaging systematics (due to galactic dust, imaging depth, and a host of other reasons) are corrected for by systematic weights at the catalog level. The imaging template-fitting techniques used (based on random forest or neural nets) damp large-scale angular modes: we measure this ``angular integral constraint" effect in mock realisations, and remove it from the power-spectrum measurements. Finally, the radial selection function imprinted in the random catalog is directly inferred from the observed data, resulting in a ``radial integral constraint" which is similarly estimated from mocks and corrected at the power spectrum level. The power spectrum covariance matrix is estimated from a set of 1000 fast approximate mocks (\texttt{EZmocks}, \cite{KP3s8-Zhao}) and rescaled to make the mock-based covariance matrix of the two-point correlation function match the semi-empirical covariance prediction obtained from the observed data (\textsc{RascalC}, \cite{KP4s7-Rashkovetskyi}). The details of this battery of validation tests are presented in \cite{DESI2024.II.KP3} and references therein. Based on these tests, we only use the monopole and quadrupole in our cosmological analysis, and restrict the full-shape analysis to the wavenumber range $0.02<k/\hmpcinv<0.20$ \cite{DESI2024.V.KP5}, with a binning width of $\Delta k = 0.005 \; \hmpcinv$.

\subsubsection{DESI full-shape blinding}
\label{sec:blinding}

An essential part of our analysis framework was ``blinding" the results during the period where data-selection and analysis choices were being made, to avoid the risk of confirmation bias.  The blinding process has two components: blinding of the BAO, and blinding of the redshift-space distortions. The blinding procedure is performed at the catalogue level, and was applied consistently to both the BAO-only analysis \cite{DESI2024.VI.KP7A} as well as the full-shape analysis in this paper. The BAO aspect of the blinding procedure follows the work of \cite{Brieden:2020}, where the redshifts of the observed galaxies are modified so that they imprint a shift in the anisotropic position of the BAO peak. The redshift-space distortion aspect of blinding, designed by \cite{Brieden:2020} to render the cosmological information about the growth of structure impervious to confirmation bias, is achieved by applying a shift in the growth rate $f$.  A full description of the blinding technique, and how it has been tailored to DESI needs, can be found in \cite{KP3s9-Andrade}.

\subsection{External data}
\label{sec:ext}

We now describe the external datasets we combine with the  DESI (FS+BAO) measurements. These choices largely follow the DESI DR1 BAO analysis \cite{DESI2024.VI.KP7A}, with the important addition of angular clustering and lensing data from the Dark Energy Survey.

We adopt the cosmic microwave background (CMB) data from the official \emph{Planck} (2018) PR3 release \cite{Planck-2018-likelihoods}.
We use as our baseline the temperature (TT) and polarisation (EE) auto-spectra, plus their cross-spectra (TE), as incorporated in the \texttt{simall}, \texttt{Commander} (for multipoles $\ell<30$) and \texttt{plik} (for $\ell\geq30$) likelihoods. As part of our robustness tests for constraints on the neutrino mass, we also alternatively consider two independent analyses of the latest \emph{Planck} PR4 data release: the high-$\ell$ \texttt{CamSpec} likelihood \cite{Efstathiou:2021,Rosenberg:2022}, and the \texttt{LoLLiPoP} (low-$\ell$) and  \texttt{HiLLiPoP} (high-$\ell$) likelihoods \cite{Tristram:2021,Tristram:2023}. We complement the CMB likelihood with the information from the reconstruction of the lensing power spectrum as measured using the connected 4-point function  of the CMB temperature and polarisation. We adopt data from the combination of \texttt{NPIPE} PR4 \Planck\ CMB lensing reconstruction \cite{Carron:2022} and the Data Release 6 of the Atacama Cosmology Telescope (ACT) \cite{Madhavacheril:ACT-DR6,Qu:2023,MacCrann:2023}.\footnote{The likelihood is available from \url{https://github.com/ACTCollaboration/act_dr6_lenslike}}  In what follows, we will denote results obtained using temperature and polarisation information from \Planck, and CMB lensing information from the \planckact\ combination, simply as ``CMB". Where necessary, we will explicitly label results that do not use CMB lensing reconstruction as ``CMB-nl".\footnote{For clarification, the TT, EE and TE power spectra always include the effect of gravitational lensing; here we emphasise that our fiducial CMB dataset additionally includes the CMB lensing reconstruction, while the CMB-nl version does not.} 

In the analyses that do not include the CMB information, we also add the prior on the physical baryon density, $\Ob h^2$, coming from Big Bang Nucleosynthesis (BBN). The theoretical BBN prediction for the abundances of light elements, especially deuterium (D) and Helium ($^4$He), is sensitive to the baryon density. Measurements of these abundances therefore lead to a constraint on the baryon density, but one which depends on details of the theoretical framework, particularly the crucial input of nuclear interaction cross-sections. As we did in our BAO-only paper \cite{DESI2024.VI.KP7A}, we adopt a recent analysis \cite{Schoeneberg:2024} that makes use of the new \texttt{PRyMordial} code \cite{Burns:2024} to recompute the predictions while marginalising over uncertainties in the reaction rates. 
We adopt the joint constraint on $\Ob h^2$ and the number of relativistic species $\Neff$, and fix the latter parameter to its fiducial value of 3.044 in models where we are not allowing for the presence of additional light relics.\footnote{To be precise, the joint constraint on $\Ob h^2$ and $\Neff$ has the respective mean values $(0.02196, 3.034)$, and the corresponding covariance (\url{https://tinyurl.com/29vzc592})
\begin{equation*}
\mathbf{C} = 
\begin{bmatrix}
4.03112260\times 10^{-7} & 7.30390042\times 10^{-5}\\
7.30390042\times 10^{-5} & 4.52831584\times 10^{-2}
\end{bmatrix}.
\end{equation*}
}

We also add information from type Ia supernovae (SN~Ia), which serve as standardisable candles offering an alternative way to measure the expansion history of the universe.  
Here we utilise the same three SN~Ia datasets that we studied in the DESI DR1 BAO paper; these are the largest compilations of supernova data that have been consistently reduced and analysed. The first SN~Ia dataset we consider is the 
PantheonPlus\footnote{We denote the originally named Pantheon+ dataset as PantheonPlus in order to avoid ambiguities with the `+' symbol used to denote the combinations of datasets.} compilation \cite{Scolnic:2021amr}, with 1550 spectroscopically-confirmed SN~Ia in the redshift range $0.001<z<2.26$, where we use the public likelihood from \cite{Brout:2022}. 
The second SN~Ia dataset that we adopt is the Union 3 compilation \cite{Rubin:2023}, containing 2087 SN~Ia in the redshift range  $0.01<z<2.26$, 1363 of which are in common with PantheonPlus, and which uses a likelihood analysis and treatment of statistical and systematic errors based on Bayesian hierarchical modelling. 
The third SN~Ia dataset is the
Year 5 supernova analysis from the Dark Energy Survey (henceforth ``DES-SN5YR"). This analysis starts with a homogeneously-selected sample of 1635 photometrically-classified SN~Ia with redshifts $0.1<z<1.3$. This is complemented by 194 low-redshift SN~Ia (which are in common with the PantheonPlus sample \cite{Scolnic:2021amr}) spanning $0.025<z<0.1$. We include all three SN~Ia datasets in our analysis; however, in certain cases where there is no meaningful dependence of the result on the choice of SN~Ia data, we only adopt one of the three datasets to avoid unnecessary redundancy.  

Additionally, we consider external information from the combination of angular galaxy clustering and weak gravitational lensing -- the so-called ``$3\times 2$-pt" datavector that consists of three two-point correlation functions (galaxy-galaxy, galaxy-shear, and shear-shear). We use results from the Dark Energy Survey Year-3 (DESY3) analysis \cite{DES:2021wwk}, which is based on observations of about 100 million source galaxies, and about 10 million lens galaxies in the fiducial \texttt{MagLim} sample, over a footprint of 4143 square degrees. The DESY3 analysis employs photometric redshifts to identify the ``source" galaxies and divide them into four tomographic bins, and the ``lens" galaxies that are subdivided into six redshift bins, although the two highest-redshift lens samples are not used in the fiducial analysis. The positions of lens galaxies are used to compute the galaxy angular clustering signal (i.e.\ galaxy-galaxy correlations); the shear of source galaxies is used to measure cosmic shear (i.e.\ shear-shear correlations); and finally the shear of source galaxies correlated with positions of the lens galaxies gives the shear-galaxy correlations.  The dataset also includes the ratio of galaxy-shear correlations at small scales in the so-called shear-ratio data vector.  The DES $3\times 2$-pt analysis mitigates information from scale-dependent bias, baryon feedback effects and nonlinearities, which are challenging to model sufficiently accurately, using a combination of scale cuts and theoretical modeling using halofit \cite{Smith:2002dz,Takahashi:2012em}. The analysis also marginalises over nuisance parameters that encode imperfect knowledge of certain astrophysical effects (such as galaxy biases, photo-$z$ distribution shifts, intrinsic galaxy alignments and multiplicative shear biases in each source tomographic bin).  For modified gravity, when we use  DES $3\times 2$-pt data, we employ a different likelihood tailored to modified-gravity (MG) analysis with similar conservative scale cuts as imposed in the DESY3 MG study \cite{DES:2022ccp}. 
We assume that the DESY3 ($3\times 2$-pt) data are uncorrelated with DESI (FS+BAO). 

In addition to the DESY3 $3\times 2$-pt data, we also make use of the so-called ``$6\times 2$-pt" data from DES Y3, which complement the galaxy clustering, cosmic shear, and galaxy-galaxy lensing with the information from the CMB lensing. Specifically, the  $6\times 2$-pt datavector extends the $3\times 2$-pt one by further adding 
the galaxy-CMB lensing, shear-CMB lensing and CMB lensing-CMB lensing two-point correlation functions. We adopt the data vector from the DESY3 $6 \times 2$-pt analysis~\cite{DES:2022urg} which uses CMB lensing data from \emph{Planck} and around $1,800$ square degrees of the South Pole Telescope (SPT)~\cite{SPT} footprint. We use the same modelling and scale cut choices as the DESY3 analyses. When combining the DESY3 $6 \times 2$-pt likelihood with the CMB, we use the CMB data without lensing (CMB-nl) in order to avoid double-counting the CMB lensing information.

\subsection{Modeling and likelihood}
\label{sec:modeling}

Having described the DESI and external datasets, we now discuss the likelihood pipeline, including the parameter space that we constrain and other details of cosmological inference. We start with a brief overview of how we theoretically model the power spectra in our ``\FM" approach.

\subsubsection{Full-shape modeling approach}
\label{sec:fs_modeling}

We use a perturbation-theory approach to full-shape clustering analysis (referred to as \FM\ here and in companion papers). [Ref.~\cite{DESI2024.V.KP5} also describes another approach called ShapeFit \cite{BriedenShapeFit}, which we use for testing and validation of our pipeline, but not for producing the cosmological results in this paper.]

The idea behind \FM\  is to directly fit a model to the full-shape power spectrum multipoles \cite{Ivanov:2019pdj,DAmico:2019fhj,Chen:2021wdi}. In this approach, we model the linear matter power spectrum using a set of cosmological parameters (see \cref{sec:inference}), and complement it with a set of nuisance parameters that describe the anisotropic power spectrum in the mildly nonlinear regime as well as various astrophysical or instrumental systematic uncertainties (e.g.\ galaxy bias).  

Our theoretical model for two-point galaxy clustering is built around cosmological perturbation theory (PT; \cite{Bernardeau02,Ivanov:2022mrd}).  Within PT, the growth of structure is treated systematically by expanding order-by-order in the amplitude of the initial fluctuations, with nonlinearities at small scales encoded using a series of ``counterterms'' that are constrained by the symmetries of the equations of motion (often known as ``effective-field theory techniques'' \cite{Ivanov:2022mrd}).  Biased tracers of large-scale structure, like galaxies or neutral hydrogen in the intergalactic medium, are treated in a consistent manner by identifying the contributions to their clustering signature allowed by fundamental symmetries at each order in PT \cite{Desjacques:2016bnm}.  Currently the redshift-space power spectrum of galaxies can be modeled with accuracy well beyond the expected statistical uncertainty in DESI, with the models being extensively tested against simulations \cite{Nishimichi2020,Chen21}, compared to each other, and tested on earlier surveys such as BOSS and eBOSS.  We have tested and compared several perturbation theory codes, and chosen to use the Eulerian PT implementation in \texttt{velocileptors} \cite{Chen20} as our default, though the results should be indistinguishable using other codes (see further discussion below).

Each of the theory codes that we employ computes the 1-loop PT predictions for the power spectrum multipoles, including mode-coupling due to quasi-linear evolution, scale-dependent bias and redshift-space distortions.  The framework includes the aforementioned counterterms that describe the impact of small-scale physics on the observed clustering, the stochastic terms that describe the shot noise and fingers of god in this formalism, and infrared resummation that describes the broadening of the BAO peak due to large-scale flows.  The models have been extensively developed and are described in some detail in \cite{KP5s2-Maus,KP5s3-Noriega,KP5s4-Lai,KP5s5-Ramirez} with references therein to the original literature.  They are compared to each other, and to a series of simulations, in \cite{KP5s1-Maus,DESI2024.V.KP5}.

A particular advantage of perturbative models of large-scale structure is they rely on a minimal set of theoretical assumptions to consistently model a wide range of clustering data.  They can thus be relied upon for robust inference.  A drawback of this approach is that these models tend to require a large number of free parameters.  If a signal in the data can be explained by a complex bias model rather than, or in addition to, changes in the underlying cosmology, the models will explore this possibility in the fits.  The majority of the cosmological information then originates from scales that are almost linear and protected by fundamental symmetries.  Unfortunately, some of the ``nuisance parameters" are partially degenerate with cosmological parameters influencing the shape of the linear theory power spectrum (e.g.\ $\Om$ and $h$). This degeneracy can cause a 
parameter ``projection effect", where the peak of the marginalised posterior is offset from the global maximum of the posterior (maximum \textit{a posteriori} value, MAP). Of particular concern are degeneracies with non-linear bias parameters, stochastic terms and counterterms that describe the impact of poorly-understood, small-scale physics on the observed clustering.  The origin and impact of these effects is discussed in detail in supporting papers to this work \cite{KP5s2-Maus,DESI2024.V.KP5} as well as in \cite{Hadzhiyska:2023wae} for example.  We do not show any results that are subject to significant projection effects, but we illustrate how such effects can occur in \cref{sec:projection}.

In addition to systematics related to theoretical modeling, we quantify several further potential systematic effects using mock catalogues. These mocks are built from the \texttt{AbacusSummit} suite of simulations \cite{AbacusSummit,abacusnbody} with a galaxy-halo connection prescription based on halo occupation distribution (HOD) models calibrated on the DESI Early Data Release \cite{DESI2023b.KP1.EDR} which are described in \cite{EDR_HOD_LRGQSO2023, EDR_HOD_ELG2023, EDR_BGS_ABACUS}.
We have identified and studied seven sources of systematic effects that could bias our cosmological constraints: i) theoretical modeling mentioned above \cite{KP5s2-Maus}, ii) description of the galaxy-halo connection \cite{KP5s7-Findlay}, iii) assumptions related to the fiducial cosmology \cite{KP5s8-Findlay}, iv) imaging systematics due to inhomogeneities in the target selection \cite{KP5s6-Zhao}, v) fibre assignment incompleteness \cite{KP3s5-Pinon,KP3s6-Bianchi,KP3s7-Lasker}, vi) spectroscopic redshift uncertainties and catastrophic redshift errors \cite{KP3s4-Yu,KP3s3-Krolewski}, and vii) covariance matrix estimation \cite{KP4s6-Forero-Sanchez,KP4s7-Rashkovetskyi,KP4s8-Alves}. Of these, the two most dominant sources of systematic effects are uncertainties associated with the imaging systematics, and the galaxy-halo connection as described by the HOD formalism. To help alleviate imaging systematics we adopt an additional nuisance parameter; more details can be found in \cite{KP5s6-Zhao}. In order to propagate the systematic errors from the HOD to the constraints on cosmological parameters, we estimate the effects on the power spectrum as described in \cite{KP5s7-Findlay}. These systematic contributions are directly added to the statistical power spectrum covariance matrix introduced in \cref{sec:FS_meas}. Detailed quantification of the systematic error budget for cosmological parameters in the \lcdm\ model is presented by \cite{DESI2024.V.KP5}.

In addition to the power spectrum measurements, we include distance-scale information from the post-reconstruction correlation function in the region around the BAO peak. The DESI DR1 BAO measurements are described in detail in \cite{DESI2024.III.KP4}, and have already been used in the DR1 BAO cosmological analysis \cite{DESI2024.VI.KP7A}.\footnote{For the analysis in the present paper, we rerun the BAO measurements with the most up-to-date catalogs as described in Appendix B of \cite{DESI2024.II.KP3}.} The joint covariance between the power spectrum and the post-reconstruction BAO measurements is estimated from the 1000 \texttt{EZmocks}. The post-reconstruction-BAO part of the resulting covariance matrix is replaced with that estimated from the BAO fits to the data. To this covariance we further add systematic contributions to both the power spectrum and post-reconstruction BAO, as summarised above for the power spectrum, and as detailed in \cite{DESI2024.III.KP4} for the BAO measurements. The full pipeline for DESI DR1 modeling analysis and the cosmological constraints in \lcdm\ from full-shape alone, and full-shape combined with BAO, are presented in \cite{DESI2024.V.KP5}.

\subsubsection{Likelihood and priors}
\label{sec:likelihood}

The combined DESI full-shape and BAO likelihood is implemented using the theoretical modeling summarised in \cref{sec:fs_modeling} and fully described by \cite{DESI2024.V.KP5}. The key data input to the likelihood are the measurements of the monopole and quadrupole of the power spectrum, restricted to scales $0.02 < k / (\hinvmpc) < 0.2$. These measurements are performed for each of the six data samples, with corresponding six redshift bins  that are listed in \cref{sec:DESI_data}.
In each redshift bin, the clustering measurements are complemented by post-reconstruction BAO parameters \cite{DESI2024.III.KP4}, and we make use of the complete covariance matrix that covers the  power spectrum measurements, the post-reconstruction BAO parameters, and their mutual correlation (see \cref{sec:fs_modeling}). The measurements in the six redshift bins are considered independent (see Sec.\ 2.3.2 of \cite{DESI2024.VI.KP7A} for a justification and quantification of inter-bin correlations), and their log-likelihoods are summed to compute the total likelihood. We combine this likelihood with the \lya BAO likelihood \cite{DESI2024.IV.KP6}, as we did in our DR1 BAO analysis \cite{DESI2024.VI.KP7A}. 

The procedure that we just described comprises our DESI (FS+BAO) likelihood. When we combine our results with external data from Dark Energy Survey clustering analyses or from the CMB, we adopt the likelihoods provided by these respective collaborations. For type Ia supernova datasets, we assume that the likelihood in the data (distances to individual supernovae) is Gaussian; this assumption has been validated to some extent with simulations (e.g.\ \cite{DES:2024hip}).

We next describe the non-cosmological ``nuisance" parameters in our analysis.
To enable the modeling of redshift-space distortions in our likelihood we adopt the Eulerian PT model in \texttt{velocileptors} \cite{KP5s2-Maus}.  The Eulerian \texttt{velocileptors} redshift-space distortion model produces posteriors that are nearly indistinguishable from those of the Lagrangian PT given DESI DR1 precision, whilst being significantly faster (a single-model evaluation for six redshift bins takes $\simeq 0.5 \; \mathrm{s}$, computing 1-loop terms once and rescaling them to each redshift). % \alej{I thought it computes the 1-loop once and then rescale with $D^4$}. 

This model also includes the scale-dependent impact of massive neutrinos on the growth rate.\footnote{\url{https://github.com/sfschen/velocileptors/blob/master/velocileptors/EPT/ept_fullresum_varyDz\_nu\_fftw.py}} 
We describe galaxy bias with three (Lagrangian-bias) parameters per redshift bin: $b_{1}$, $b_{2}$, and $b_{s}$. The third-order bias parameter, $b_{3}$, is expected to be small and is degenerate with the other nuisance parameters, so we set it to zero; see \cite{KP5s2-Maus} for tests validating this choice. In practice, we sample and impose priors on $(1 + b_{1})\sigma_8$, $b_{2}\sigma_8^2$, and $b_{s}\sigma_8^2$ (with each $\sigma_8$ evaluated at the effective redshift of the corresponding bin), as the data are sensitive to these combinations. Next, we include stochastic parameters $\mathrm{SN}_0$ and $\mathrm{SN}_2$ which marginalise over small-scale physics (halo exclusion effects, conformity, and virialisation), and enter additively to the anisotropic power spectrum as $\mathrm{SN}_0$ and $\mathrm{SN}_2k^2\mu^2$ (where $\mu$ is the cosine between the wavenumber $\mathbf{k}$ and the line-of-sight to the galaxy pair); see Eq.~(3.6) in \cite{KP5s2-Maus}. The priors on the stochastic parameters are Gaussian with mean zero; the widths of these Gaussians are given in \cref{tab:priors}, and are further scaled by the estimated shot noise (for $\mathrm{SN}_0$), or by the product of the shot noise with a typical velocity variance and the satellite fraction (for $\mathrm{SN}_2$) \cite{KP5s2-Maus}. To describe the modeling uncertainties associated with non-linear structure formation, beyond the cutoff scale adopted in one-loop integrals of the perturbation theory model, we include two ``counter-term" parameters, $\alpha_0$ and $\alpha_2$ corresponding, respectively, to the monopole and the quadrupole. We give each of these parameters a Gaussian prior centered at zero with a standard deviation of $12.5$; this prior width is chosen to correspond to the value at which the counter term corrections represent 50\% of the value of the linear power spectrum contribution at the maximum wavenumber $k_{\rm max}=0.2\,\hinvmpc$ used in our fits. As a sanity check, we tested the likelihood using the PT code \texttt{FOLPS} \cite{KP5s3-Noriega,Noriega:2022nhf} as an alternative to \texttt{velocileptors}, finding nearly identical results.  The bottom part of \cref{tab:priors} summarises the nuisance parameters and their priors.  Overall, we adopt three bias, two counterterm, and two stochasticity parameters in each of the six redshift bins, resulting in the grand total of 42 nuisance parameters.

Our fiducial constraints, which we refer to as DESI (FS+BAO), are based on this combined power spectrum and post-reconstruction BAO likelihood. Whenever we do \textit{not} add the CMB data to DESI, we include two non-trivial external priors as a default. First, we combine DESI with the external constraint on the physical baryon density $\Ob h^2$ that comes from  measurements of the primordial deuterium abundance and helium fraction  interpreted in the standard model of Big Bang Nucleosynthesis (BBN) \cite{Schoeneberg:2024}; the model predictions were generated using the \texttt{PRyMordial} code \cite{Burns:2024}; see \cref{sec:ext} for more details. Second, we add a weak Gaussian prior on the spectral index $\ns$, centered at the \emph{Planck} mean value $\ns=0.9649$ and with a width, $\sigma(\ns) = 0.042$, chosen to be 10 times wider than the posterior width obtained from \emph{Planck} temperature, polarisation and lensing spectra \cite{Planck-2018-cosmology}.\footnote{The loose $\ns$ prior was originally imposed to stabilise the results of the ShapeFit analyses used in our tests \cite{DESI2024.V.KP5}. This prior has a small effect on the cosmological results from the Full-Modeling analyses presented in this paper. In cases when we add this prior, we still impose hard prior cutoffs $\ns\in [0.8, 1.2]$.} The loose $\ns$ prior to which we refer to as $\nsten$, is therefore implemented as
%%%%%%%%%%%%%%%%%%%%%%%%%%%%%%%%
\begin{equation}
\ns \sim \mathcal{N}(0.9649, 0.042^2)
\qquad (\nsten\,\,\, \mbox{prior}),
\label{eq:DESI_priors}
\end{equation}
%%%%%%%%%%%%%%%%%%%%%%%%%%%%%%%%
where $\mathcal{N}(x, \sigma^2)$ refers to the Gaussian normal distribution with mean $x$ and standard deviation $\sigma$. When we combine DESI data with the CMB, we do not apply the BBN and $\nsten$ priors, as the CMB already tightly constrains these two parameters.

\subsubsection{Cosmological inference}
\label{sec:inference}

Our inference procedure largely follows that described in the DESI DR1 BAO paper \cite{DESI2024.VI.KP7A}; the main difference is that we now marginalise over many more nuisance parameters which are specific to the full-shape analysis.  We use the cosmological inference code \texttt{cobaya} \cite{Torrado:2019,Torrado:2021}, to which we incorporate the PantheonPlus, Union3 and DES-SN5YR SN~Ia likelihoods, as well as our DESI likelihood.
We use CMB likelihoods based on public packages that are either included in the public \texttt{cobaya} version or available directly from the respective teams. Within \texttt{cobaya} we use the Boltzmann code \texttt{CAMB}~\cite{LewisCAMB:2000, HowlettCAMB:2012} to produce model power spectra as a function of parameters. For modified-gravity analyses we employ the code \texttt{ISiTGR}~\cite{Dossett:2011tn,Garcia-Quintero:2019xal} which is based on \texttt{CAMB} and also called within \texttt{cobaya}. When including the CMB data (the combined \Planck+ACT lensing likelihood) we use higher precision settings as recommended by ACT. 
We perform Bayesian inference using the Metropolis-Hastings MCMC sampler \cite{LewisMCMC:2002, LewisMCMC:2013} in \texttt{cobaya}, running four chains in parallel. We use \texttt{getdist}\footnote{\url{https://github.com/cmbant/getdist}}~\cite{Lewis:2019xzd} to derive the constraints presented in this paper. We occasionally wish to calculate $\Delta \chi_\mathrm{MAP}^2 \equiv -2 \Delta \log{\mathcal{L}}$, defined to represent the difference (times $-2$) of log-posteriors $\log{\mathcal{L}}$ at the maximum posteriori (MAP) parameter-space points. Such MAP poi are estimated with the \texttt{iminuit} \cite{iminuit,James:1975dr} algorithm, as implemented in \texttt{cobaya}, starting from the points with maximum log-posterior found in the MCMC chains. 
More details about the code settings and extraction of the cosmological-parameter values is provided in Sec 2.5 of the DESI DR1 BAO paper \cite{DESI2024.VI.KP7A}.

\begin{table}[t] 
    \centering
    \begin{tabular}{|lllll|}
    \hline
    data or model & parameter & default & prior & comment\\  
    \hline 
    \textbf{DESI (\lcdm)}
     & $\Ob h^2$ &---& $\mathcal{N}(0.02218, 0.00055^2)$ &---  \\
    & $\ns$ &---& $\mathcal{N}(0.9649, 0.042^2)$ &\emph{Planck} 10$\sigma$\\  
    & $\Ocdm h^2$ &---& $\mathcal{U}[0.001, 0.99]$ &---\\
    & $\ln(10^{10} A_\mathrm{s})$ &---& $\mathcal{U}[1.61, 3.91]$ &---\\    
    \hline     
    \textbf{CMB (\lcdm)} 
    & $100 \theta_{\mathrm{MC}}$ &---& $\mathcal{U}[0.5, 10]$ &replaces $H_0$\\
    & $\tau$ & 0.0544 & $\mathcal{U}[0.01, 0.8]$ &---\\
    & $\Ob h^2$ &---& $\mathcal{U}[0.005, 0.1]$ & ---\\ 
    & $\ns$ &---& $\mathcal{U}[0.8, 1.2]$ & \mbox{no 10$\sigma$ prior}\\      
    \hline 
    \textbf{Beyond \lcdm} 
    %& $\Ok$ & $0$ & $\mathcal{U}[-0.3, 0.3]$ &\\
    & $w_0$ & $-1$ & $\mathcal{U}[-3, 1]$ &---\\
    (dynamical DE) & $w_{a}$ & $0$ & $\mathcal{U}[-3, 2]$ &---\\
    \hdashline    
    (massive $\nu$) & $\sumnu \; (\eV)$ & $0.06$ & $\mathcal{U}[0, 5]$ & --- \\
    \hdashline    
    & $\mu_0$ & $0$ & $\mathcal{U}[-3, 3]$ & ---\\
    (modified gravity) & $\Sigma_0$ & $0$ & $\mathcal{U}[-3, 3]$ &--- \\
    \hline
    \textbf{nuisance (DESI)} 
    & $(1 + b_1) \sigma_8$ &  & $\mathcal{U}[0, 3]$ & each $z$-bin\\    
    & $b_2 \sigma_8^2$ &  & $\mathcal{N}[0, 5^2]$ & each $z$-bin\\    
    & $b_s \sigma_8^2$ &  & $\mathcal{N}[0, 5^2]$ & each $z$-bin\\ 
    & $\alpha_0$ &  & $\mathcal{N}[0, 12.5^2]$ & each $z$-bin$^*$\\
    & $\alpha_2$ &  & $\mathcal{N}[0, 12.5^2]$ & each $z$-bin$^*$\\
    & $\mathrm{SN}_0$ &  & \hspace{-0.4cm}$\propto\mathcal{N}[0, 2^2]$ & each $z$-bin$^*$\\
    & $\mathrm{SN}_2$ &  & \hspace{-0.4cm}$\propto\mathcal{N}[0, 5^2]$ & each $z$-bin$^*$\\
   \hline 
    \end{tabular}
    \caption{
    Parameters and priors used in our analysis. Here, $\mathcal{U}$ refers to a uniform prior in the range given, whilst $\mathcal{N}(x, \sigma^2)$ refers to the Gaussian normal distribution with mean $x$ and standard deviation $\sigma$. In addition to the flat priors on $w_0$ and $w_a$ listed in the table, we also impose the requirement $w_0+w_a<0$ in order to enforce a period of high-redshift matter domination. Similarly, an extra prior $\mu_0 < 2 \Sigma_0 + 1$ is included for modified-gravity parameters $\mu_0$ and $\Sigma_0$ (see \cref{sec:MG_constraints}).     
    Nuisance-parameter combinations listed in the second column are independently varied for each of the six tracer/redshift bins. The asterisk next to the counter-terms $\alpha_0$ and $\alpha_2$ and stochastic parameters $\mathrm{SN}_0$ and $\mathrm{SN}_2$ indicates that these parameters are marginalised over analytically. The constant of proportionality in front of $\mathrm{SN}_0$ and $\mathrm{SN}_2$ priors indicates that these priors as written are further scaled with corresponding physically motivated terms; see text for details.  We note that the BBN and $\ns$ priors are added by default in the DESI (FS+BAO) analysis, but dropped once DESI data is combined with the CMB. 
    }
    \label{tab:priors}
\end{table}

\Cref{tab:priors} summarises the cosmological parameters that are sampled over in different runs and the priors that are placed on them. For the basic DESI (FS+BAO) analysis and assuming the \lcdm\ model, we vary five cosmological parameters: the Hubble constant $H_0$, the physical densities of baryons and cold dark matter $\Ob h^2$ and $\Ocdm h^2$, and the amplitude and spectral index of the primordial density perturbations, $A_\mathrm{s}$ and $\ns$. When we add the CMB likelihood, instead of $H_0$ we vary the parameter $\theta_{\rm MC}$ which is an approximation to the acoustic angular scale $\theta_\ast$, and we add the optical depth to reionization parameter, $\tau$. In models beyond \lcdm\, we extend this basic cosmological parameter set with additional variables: in the dynamical dark-energy (\wowacdm) model we have two additional dark-energy parameters $w_0$ and $w_a$, in the massive-neutrinos model the sum of the neutrino masses $\sumnu$, and in the class of modified-gravity parametrisation we consider, we introduce additional freedom in the linearly-perturbed Einstein's equations given by parameters $\mu_0$ and $\Sigma_0$ (see \cref{sec:MG} for their definitions). Finally, we have a set of nuisance parameters which are required to describe the full-shape clustering signal.  In all, our analysis in the base \lcdm\ model includes a total of five cosmological parameters and 42 nuisance parameters.

In \lcdm, we combine the DESI (FS+BAO) analysis with the DESY3 $3 \times 2$-pt and $6 \times 2$-pt analyses at the likelihood level, because there is negligible covariance between the multipoles and the projected statistics~\cite{Taylor:2022rgy}. We rerun the  DESY3 $3 \times 2$-pt and $6 \times 2$-pt analyses with the same priors as the DESI (FS+BAO) analysis using the publicly available {\tt CosmoSIS}~\cite{Zuntz:2014csq} pipelines. For each cosmological model, we use the same modeling and scale cuts.  For these combinations we then use  {\tt CombineHarvesterFlow}\footnote{\url{https://github.com/pltaylor16/CombineHarvesterFlow}}~\cite{Taylor:2024eqc} to fit normalising flows to the DES chains and re-weight the DESI (FS+BAO) chains to compute the joint posteriors. To ensure the results are not affected by undersampling the joint high-density region~\cite{Taylor:2024eqc}, we randomly split the DES and DESI chains in half and repeat this procedure on both pairs of chains and check the results remain unchanged. 

For our modified-gravity inference, we use a DESY3 $3 \times 2$-pt likelihood that has been tailored to this specific analysis and validated against the DESY3 modified-gravity results by \cite{DES:2022ccp}. This likelihood has been included in our main pipeline using \texttt{cobaya}.

%%%%%%%%%%%%%%%%%%%%%%%%%%%%%%%%%%%%%%%%%%%%%%%%%%%%%%%%%%%%%%

\section{Dark energy constraints}
%\lcdm\, and \wowacdm\, from DESI full-shape including external data sets}
\label{sec:cosmo_constraints}
%%%%%%%%%%%%%%%%%%%%%%%%%%%%%%%%%%%%%%%%%%%%%%%%%%%%%%%%%%%%%%

We focus on two dark energy models: a spatially flat model with a cosmological constant (\lcdm), and a flat model where the dark energy equation of state is allowed to vary with time and is modeled by two parameters (\wowacdm). We choose not to study the models with non-zero spatial curvature ($\Ok$) or with a constant dark energy equation of state (\wcdm) in this paper as we did in the analysis of DESI BAO-only data \cite{DESI2024.VI.KP7A}. Instead, we focus on the two aforementioned models that are of most interest: \lcdm\ because it is the standard model of cosmology, and \wowacdm\ because of its ability to phenomenologically describe a wide variety of physical models, and because of our earlier findings that show some preference for this model over \lcdm\ \cite{DESI2024.VI.KP7A}.

\subsection{$\Lambda$CDM model}\label{subsec:LCDM-model}

We start by constraining the cosmological parameters in a flat \lcdm\ model. Here, only a single parameter describes dark energy: $\Ode\equiv 1-\Om$, where $\Om$ and $\Ode$ are respectively the total matter and dark-energy densities relative to critical. The other parameters that we vary, and their respective priors, are listed in \cref{tab:priors}.
Other than $\Om$, the parameters of most interest to us are the amplitude of mass fluctuations $\sigma_8$ (and the combination $\Seight\equiv \sigma_8 (\Om/0.3)^{0.5}$), and the Hubble constant $H_0$.

\begin{figure*}
\includegraphics[width=\linewidth]{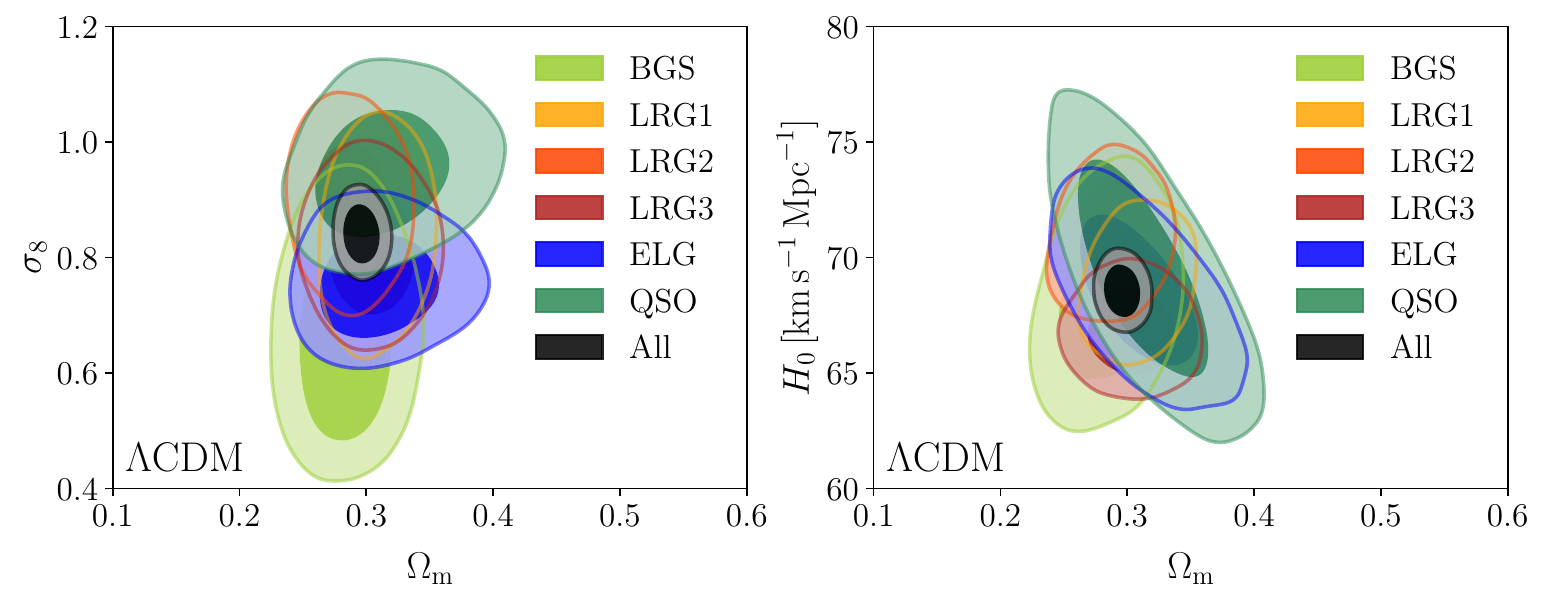}
\caption{
68\% and 95\% credible intervals in the $\Om$--$\sigma_8$ plane (left panel) and  $\Om$--$H_0$ plane (right panel) from the combined DESI full-shape and BAO analysis, assuming the \lcdm\ background.  We show the constraints from individual DESI tracers (with the BBN and loose $\ns$ priors), and the combined measurement that includes all the tracers shown and the \lya BAO data.
}
\label{fig:Om_s8_H0}
\end{figure*}

Figure \ref{fig:Om_s8_H0} shows the DESI-only constraints (combined with the BBN and loose $\ns$ priors) in the $\Om$--$\sigma_8$ plane (left panel) and $\Om$--$H_0$ plane (right panel) from the combined DESI full-shape and BAO analysis, marginalised over all other parameters. The individual contours, with their 68\% and 95\% credible intervals, show measurements from individual DESI tracers: bright galaxy survey (BGS); luminous red galaxies in three redshift bins (LRG1, LRG2 and LRG3), emission line galaxies (ELG) and quasars (QSO). The combined constraints from these tracers, including their covariance and also including the geometrical (BAO) information from the \lya\ forest, is shown by the small black contour in each panel. This figure illustrates the excellent mutual agreement between individual tracers, as well as their complementarity. The combined measurements on these three parameters from all DESI tracers are,
%%%%%%%%%%%%%%%%%%%%%%%%%%%%%%%%%%%%%%%%%%
\threeonesig[5.5cm]
{\Om &= 0.2962\pm 0.0095,}
{\sigma_8 &= 0.842\pm 0.034,}
{H_0 &= (68.56\pm 0.75)\,\kmsMpc,}
{DESI (FS+BAO)+BBN+$\nsten$, \label{eq:DESI_LCDM}}
%%%%%%%%%%%%%%%%%%%%%%%%%%%%%%%%%%%%%%%%%%%
where, recall, we add the BBN and loose $\ns$ priors to DESI-only data by default.
The addition of full-shape information therefore leads to a significant improvement in the constraints on matter density, from $\sigma(\Om)=0.015$ in the DESI (BAO) + BBN case \cite{DESI2024.VI.KP7A}, to $\sigma(\Om)=0.0095$ when the full-shape information (along with our loose $\ns$ prior) is added. This measurement is roughly comparable to, and almost entirely independent of, the constraint from the CMB alone ($\sigma(\Om)=0.0066$, for our fiducial CMB combination). 

Note that the results of \cref{eq:DESI_LCDM} include a $4$\% constraint on the amplitude of mass fluctuations, $\sigma_8$. The BAO alone are a purely geometric probe and thus insensitive to $\sigma_8$, so the present constraint comes from using the full-shape clustering measurements of DESI's tracers.  DESI's constraint on $\sigma_8$ is broadly consistent with that from other cosmological measurements, although with an error much larger than that from the CMB data alone (which give $\sigma_8 = 0.8133\pm 0.0050$).

\begin{figure*}
\includegraphics[width=14.0cm]{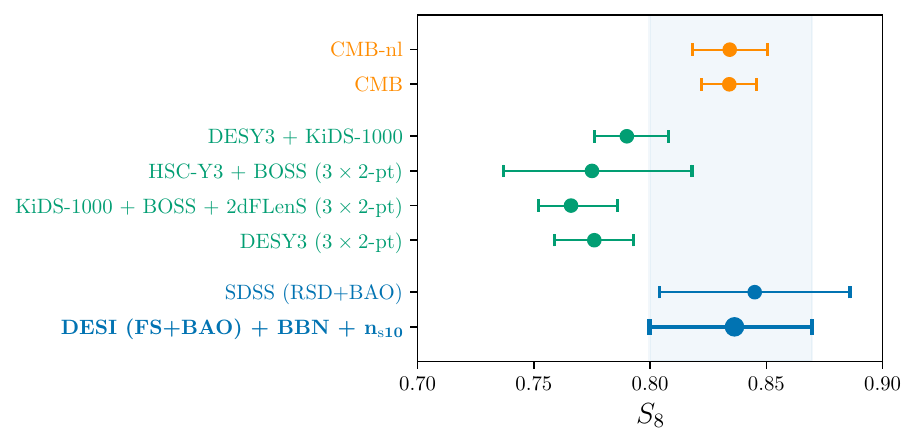}
\caption{
Constraints on the parameter $\Seight\equiv \sigma_8 (\Om/0.3)^{0.5}$, assuming the \lcdm\ background.  The whisker on the bottom (and the corresponding blue-shaded region) shows our fiducial 68\% constraint from DESI DR1 (FS+BAO), combined with the BBN and loose $\ns$ priors. The first two whiskers from the top, in orange, show the constraints from the CMB, without and with CMB lensing information. The following four whiskers, in green, show the results from weak lensing probes, while the second to bottom whisker, in blue, shows the constraints from the SDSS combination of redshift-space distortions and BAO. See text for details. 
}
\label{fig:S8}
\end{figure*}

We next study constraints on the derived parameter $\Seight\equiv \sigma_8 (\Om/0.3)^{0.5}$. This parameter combination (unlike $\sigma_8$ by itself) is accurately measured by weak gravitational lensing probes of the large-scale structure, and is thus a good meeting point to compare results from galaxy clustering (and CMB and SNIa) surveys with those from  weak gravitational lensing.  \cref{fig:S8} shows the marginalised 68\% constraints on $\Seight$. The bottom whisker in the Figure  shows our fiducial 68\% constraint from DESI DR1 FS+BAO (and the shaded region shows the same)
%%%%%%%%%%%%%%%
\oneonesig[5.5cm]{\Seight= 0.836\pm 0.035}{DESI (FS+BAO)+BBN+$\nsten$}{. \label{eq:S8_DESI}} 
%%%%%%%%%%%%%%%
Next, the top two whiskers in  \cref{fig:S8} show constraints from the CMB, without and with CMB lensing information. The following four rows display results from the combined shear analysis from the Dark Energy Survey and Kilo-Degree Survey (DESY3+KiDS-1000; \cite{Kilo-DegreeSurvey:2023gfr}); constraints from shear measured in the Hyper Suprime-Cam Year-3 data combined with galaxy clustering from the Baryon Oscillation Spectroscopic Survey (HSC Y3 + BOSS; \cite{2023PhRvD.108l3521S}); constraints from shear measured in the photometric KiDS-1000 survey combined with spectroscopic galaxy clustering in BOSS and the 2-degree Field Lensing Survey (2dFLenS) (KiDS-1000+BOSS+2dFLenS; \cite{Heymans:2020gsg}); and the combined shear and clustering $3\times 2$-pt analysis from the DESY3 data \cite{DES:2021wwk}. The second-to-bottom whisker shows the constraints from the SDSS RSD+BAO analysis \cite{eBOSS:2020yzd} that used a full-shape analysis that was substantially different in detail to our \FM. We observe an excellent agreement between DESI DR1 (FS+BAO) and the CMB, both of which are slightly higher than the values inferred from the weak lensing surveys. There remains some modest difference between our constraint in \cref{eq:S8_DESI} and that from the lensing probes; for example, the DESY3 measurement is $\Seight = 0.776\pm  0.017$ \cite{DES:2021wwk}. We do not discuss these discrepancies further in this paper. We also note the excellent agreement between our results and the SDSS combined of redshift-space distortion and BAO analysis \cite{eBOSS:2020yzd}, which found $\Seight= 0.845\pm 0.041$. Finally, it is interesting to compare our constraints to those resulting from \textit{cross-correlating} a spectroscopically-calibrated LRG sample selected from the DESI Legacy imaging survey \cite{LS.Overview.Dey.2019} with lensing probed by galaxies or the CMB.  Cross-correlating with CMB lensing gives $\Seight=0.775\pm 0.02$ \cite{Sailer:2024coh,ACT:2024okh}, lower than but consistent with our result (see also \cite{ACT:2023oei} for a related analysis that gives similarly consistent constraints).  Cross correlating the DESI galaxy positions with DES galaxy shapes gives $\Seight=0.850^{+0.042}_{-0.050}$ \cite{Chen24}, again consistent with our result.

\begin{figure*}
\includegraphics[width=14.0cm]{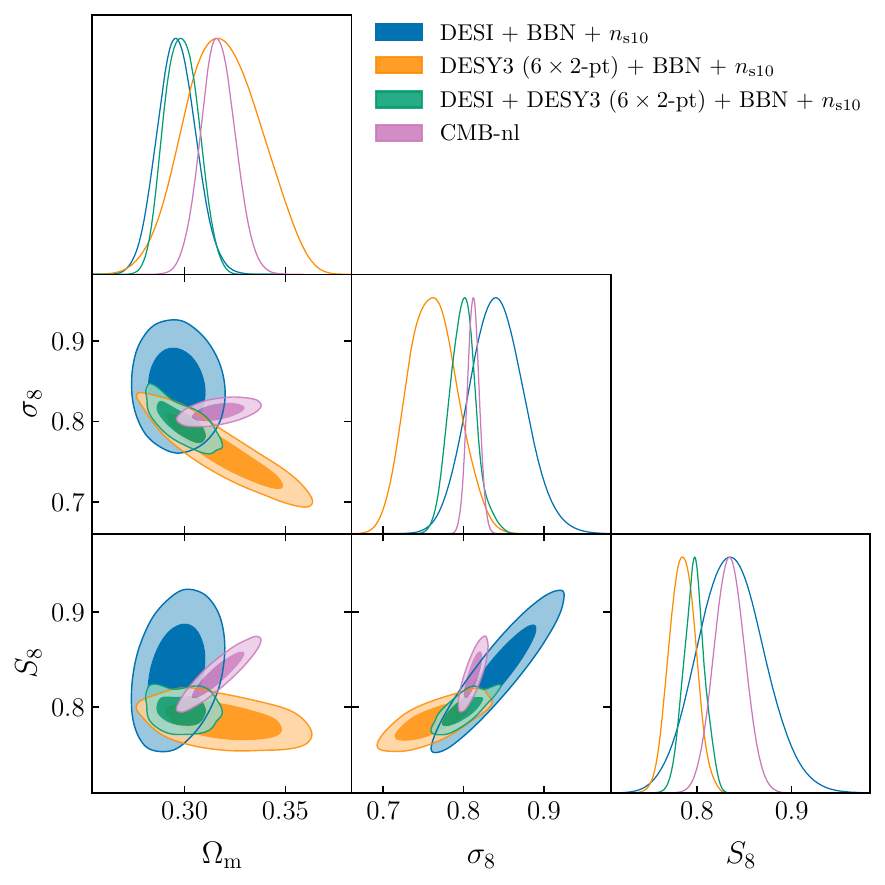}
\caption{
Projected constraints on $\Om$, $\sigma_8$, and $S_8$ in the \lcdm\ model, with 68\% and 95\% credible intervals shown in each case. The blue contours display the DESI (FS+BAO) constraints (with the BBN and loose $\ns$ priors). The orange contours show constraints from the DESY3 ($6\times 2$-pt) analysis which combines galaxy clustering, cosmic shear and CMB lensing. The green contours show the combination of DESI and DESY3 ($6\times 2$-pt) data. For comparison we also show the CMB temperature and polarisation constraints without the lensing reconstruction as purple contours. 
}
\label{fig:Om_s8_S8_LCDM}
\end{figure*}

\begin{figure*}
\centering
\includegraphics[width=10.0cm]{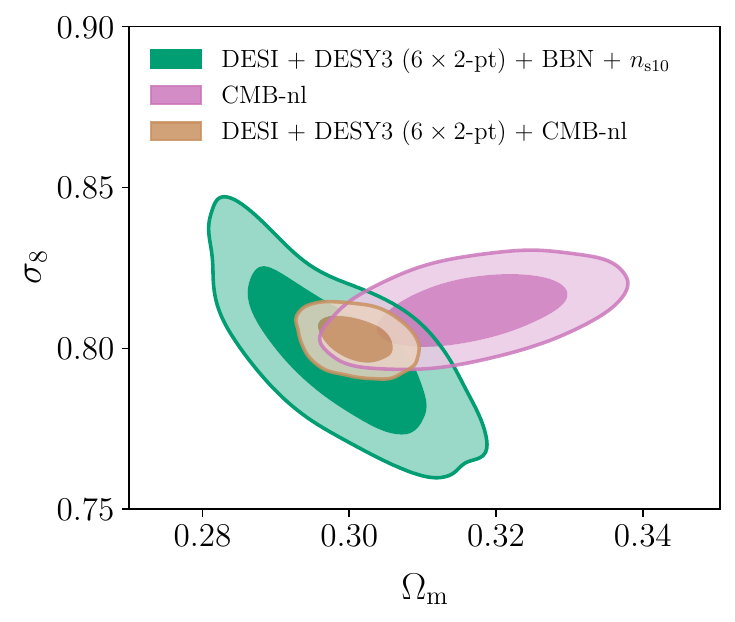}
\caption{Constraints projected to the $\Om$--$\sigma_8$ plane in the \lcdm\ model. The green contour shows constraints from the DESI full shape and BAO analysis, combined with the BBN and $\nsten$ priors, and further complemented with the DESY3 ($6\times 2$-pt) data. The pink contour shows the CMB without lensing reconstruction. The brown contour shows the combination of the two, that is, DESI combined with DESY3 ($6\times 2$-pt) and CMB-nl.}
\label{fig:Om_s8_LCDM}
\end{figure*}

%%%%%%% BEGIN TABLE %%%%%%%%%%%%%%%%%%%
\begin{table}
\centering
\resizebox{\columnwidth}{!}{%
    \small 
\setcellgapes{3pt}\makegapedcells  % for makecell command below     
\renewcommand{\arraystretch}{2.1} % row spacing
    \begin{tabular}{lcccccc}
    \toprule
    \midrule
     model/dataset & $\Om$ & $\sigma_8$ & $S_8$ & \makecell[c]{$H_0$\\[0.1cm] [${\rm km/s/Mpc}$]} & $w_0$& $w_a$\\[0.1cm]
    \midrule\midrule
    {\bf Flat} $\boldsymbol{\Lambda}${\bf CDM} &&&&\\[-0.0cm]
   \makecell[l]{DESI (FS+BAO)\\ \quad+ BBN+$\nsten$} & $0.2962\pm 0.0095$ & $0.842\pm 0.034$ & $0.836\pm 0.035$ & $68.56\pm 0.75$ &---&---\\[0.2cm]
    \hdashline    
    DESI+CMB-nl & $0.3045\pm 0.0053$ & $0.8086\pm 0.0071$ & $0.815\pm 0.012 $ & $68.14\pm 0.40$ &---& ---\\
    DESI+CMB & $0.3056\pm 0.0049$ & $0.8121\pm 0.0053$ & $0.8196\pm 0.0090$ & $68.07\pm 0.38$ &---& ---\\
    \hdashline\\[-0.7cm]    
    \makecell[l]{DESI+DESY3 ($3 \times 2$-pt)\\ \quad+ BBN+$\nsten$}   & 
    $0.2980\pm 0.0070$ & 
    $0.807^{+0.016}_{-0.020}$ & 
    $0.804^{+0.011}_{-0.015}$ & 
    $68.67^{+0.69}_{-0.77}$  & --- & ---\\[0.2cm]
    \makecell[l]{DESI+DESY3 ($6 \times 2$-pt)\\ \quad+ BBN+$\nsten$}   & $0.2986\pm 0.081$ & 
    $0.799\pm 0.016$ & 
    $0.797\pm 0.011$ & 
    $68.66^{+0.63}_{-0.73}$  & --- & ---\\[0.2cm]
    \hdashline\\[-0.7cm]
    \makecell[l]{DESI+DESY3 ($6 \times 2$-pt) \\\quad + CMB-nl }  & 
    $0.3009\pm 0.0034$ & 
    $0.8028^{+0.0050}_{-0.0045}$ & 
    $0.8039\pm 0.0056$ & 
    $68.40\pm 0.27$ & --- & --- \\
    {\bf Flat} $\boldsymbol{w_0w_a}${\bf CDM} &&&&\\[-0.2cm]
    \midrule   
     DESI+PantheonPlus  & $0.3084\pm 0.0089$ & $0.820\pm 0.035$ & $0.831\pm 0.036$ & $68.4\pm 1.1$ & $ -0.875\pm 0.072$ & $-0.61^{+0.42}_{-0.36}$ \\[-0.5cm]
    \quad+ BBN+$\nsten$& ($0.3117$) & ($0.829$) & ($0.845$) & ($67.8$) & ($-0.874$)  & ($-0.48$) 
    \\
     DESI+Union3 & $0.320\pm 0.012$ & $0.805^{+0.033}_{-0.037}$ & $0.831\pm 0.036$ & $67.6\pm 1.2$ & $-0.74\pm 0.12$ & $-1.12^{+0.58}_{-0.48}$ \\[-0.5cm]
    \quad+ BBN+$\nsten$& ($0.328$) & ($0.809$) & ($0.846$) & ($66.6$) & ($-0.68$)  & ($-1.15$)
    \\[0.1cm]
    DESI+DES-SN5YR & $0.3183\pm 0.0090$ & $0.808^{+0.033}_{-0.037}$ & $0.832^{+0.034}_{-0.038}$ & $67.7\pm 1.0$ & $-0.761\pm 0.080$ & $-1.03^{+0.47}_{-0.40}$ \\[-0.5cm]
    \quad+ BBN+$\nsten$& ($0.3214$) & ($0.815$) & ($0.843$) & ($67.2$) & ($-0.759$)  & ($-0.92$) \\
    \hdashline    
    DESI+CMB & $0.3061\pm 0.0064$ & $0.8227\pm 0.0087$ &  $0.8309\pm 0.0091$ & $68.34\pm 0.67$ &$-0.858\pm 0.061$ & $-0.68^{+0.27}_{-0.23}$ \\[-0.5cm]
    \quad +PantheonPlus & ($0.3091$) & ($0.8210$) &($0.8334$)&  ($67.92$) & ($-0.847$)  & ($-0.64$) \\
    DESI+CMB  & $0.3156\pm 0.0090$ & $0.8152\pm 0.0099$ &  $0.8360\pm 0.0097$ & $67.35\pm 0.92$ &$-0.742\pm 0.096$ & $-1.02^{+0.36}_{-0.32}$ \\[-0.5cm]
    \quad+Union3 & ($0.3246$) & ($0.8073$) &($0.8397$)&  ($66.44$) & ($-0.667$)  & ($-1.20$) \\ 
    DESI+CMB  & $0.3142\pm 0.0063$ & $0.8163\pm 0.0083$ & $0.8353\pm 0.0092$ & $67.48\pm 0.62$ & $-0.761\pm 0.065$ & $-0.96^{+0.30}_{-0.26}$ \\[-0.5cm]
    \quad+DES-SN5YR & ($0.3171$) & ($0.8157$) & ($0.8386$) & ($67.11$) & ($-0.749$)  & ($-0.92$) \\           

    \midrule
    \bottomrule
    \end{tabular}
}
\caption{ 
    Cosmological parameter results from DESI DR1 (FS+BAO) data alone (labeled “DESI” in the table), and in combination with external datasets. We show results in the baseline flat \lcdm\ model and in the $(w_0, w_a)$ parameterisation of the dark energy equation of state. Constraints are quoted as the marginalised means and 68\% credible intervals in each case. For flat \wowacdm\ model, where mild projection effects are observed in some cases, we also show the best-fit (MAP) value of the parameter in parentheses just below the credible interval. In this and other tables, the shorthand notation ``CMB" is used to denote the addition of temperature and polarisation data from \Planck\, and CMB lensing data from the combination of \Planck\ and ACT, while $\nsten$ refers to the loose prior on the spectral index defined in \cref{eq:DESI_priors}.
    \vspace{0.1em}
    \label{tab:parameter_table1}
}

\end{table}
%%%%%%% END TABLE %%%%%%%%%%%%%%%%%%%

We next discuss the results of combining DESI (FS+BAO) with external probes. When we combine DESI with the CMB information we obtain,
%%%%%%%%%%%%%%%%%%%%%%%%%%%%%%%%%%%%%%%%%%
\threeonesig[5cm]
{\Om &= 0.3056\pm 0.0049,}
{\sigma_8 &= 0.8121\pm 0.0053,}
{H_0 &= (68.07\pm 0.38)\,\kmsMpc,}
{DESI (FS+BAO) + CMB. \label{eq:DESI_CMB_LCDM}}
%%%%%%%%%%%%%%%%%%%%%%%%%%%%%%%%%%%%%%%%%%%
These results are fully consistent with those from DESI (FS+BAO) alone, but have much  smaller errors: the uncertainties in the matter density and Hubble constant are roughly halved, whilst the uncertainty in $\sigma_8$ decreases by more than a factor of five. 

We also find that the combined information from DESI full-shape clustering and the BAO, when combined with the CMB, generally improves the constraints from CMB alone: the $\Om$, $H_0$, and $S_8$ errors decrease by 30\%, but  the $\sigma_8$ error increases  slightly  (by $\sim$10\%). Our tests on synthetic chains from data with no stochasticity confirm that the observed improvement in $\Om$, $H_0$, and $S_8$ precision, and lack thereof in $\sigma_8$ precision, is expected  when DESI (FS+BAO) is added to CMB alone. Hence, a mild worsening of the constraints on $\sigma_8$ when the DESI data are added is not unexpected.

Combining DESI data with that of the CMB \textit{without} CMB lensing reconstruction (that is, without the ACT+\emph{Planck} lensing likelihood) gives similar results as the full DESI+CMB fits. The only exception is the amplitude of mass fluctuations $\sigma_8$, which now has a $\sim$30\% larger error ($\sigma_8 = 0.8086\pm 0.0071$) in the DESI+CMB-nl analysis, relative to the precision in DESI+CMB. This is to be expected, given that CMB lensing measures the depth of the lensing potential at a range of redshifts ($z\sim 0.5$--$3$, very roughly), and is hence sensitive to this parameter. See  \cref{tab:parameter_table1} for more details.\footnote{Note also that we choose not to combine the SN~Ia data with that of DESI+CMB in the \lcdm\ model, as the two respective measurements disagree in the value of $\Om$ (see the discussion in Sec.~4.1 of \cite{DESI2024.VI.KP7A}). 
}

The final external dataset we consider is the combined analysis of weak gravitational lensing and galaxy angular clustering data from the Dark Energy Survey, DESY3 ($3\times 2$-pt). These lensing and galaxy clustering measurements are sensitive to the power spectrum and growth of structure in a way that is complementary to DESI (FS+BAO), with a different set of systematic errors. We also consider the $6\times 2$-pt analysis from DESY3, which adds further information from lensing of the CMB (see \cref{sec:ext} for details). The addition of the DESY3 ($6\times 2$-pt) data to DESI (FS+BAO) produces,
%%%%%%%%%%%%%%%%%%%%%%%%%%%%%%%%%%%%%%%
\threeonesig[6cm]
{\Om &= 0.2986\pm 0.0081,\\[-0.45cm]}
{\sigma_8 &= 0.799\pm 0.016,\\[-0.45cm]}
{H_0 &= (68.66^{+0.63}_{-0.73})\,\kmsMpc.}
{DESI (FS+BAO)+BBN+$\nsten$ + DESY3 ($6\times 2$-pt). \label{eq:DESI_DESY3_LCDM}}
%%%%%%%%%%%%%%%%%%%%%%%%%%%%%%%%%%%%%%%
The addition of DESY3 ($6\times 2$-pt) data therefore  improves DESI's constraint on $\sigma_8$ by about a factor of two (compare to \cref{eq:DESI_LCDM}). It also pulls the central value of $\sigma_8$ down by about one standard deviation; the same trend is observed in $S_8$, for which the measurement error is improved by nearly a factor of three by the addition of DES data (see \cref{tab:parameter_table1}). The downward pull in $\sigma_8$ and $S_8$ is unsurprising as the lensing information present in DESY3 data favors lower values of these parameters \cite{DES:2021wwk}.  We also note that the addition of the DESY3 ($6\times 2$-pt) data does not appreciably change the precision of DESI's constraints on $\Om$ and $H_0$. Moreover, the constraints and trends remain similar if we replace the DESY3 ($6\times 2$-pt) by the DESY3 ($3\times 2$-pt) data in these tests (see \cref{tab:parameter_table1}), but the $\sim$20\% improvement in the $\sigma_8$ and $S_8$ errors when going from $3\times 2$-pt to $6\times 2$-pt data indicates that the CMB lensing information present in the latter DESY3 analysis is significant. Note finally that the uncertainty in $\Om$ \textit{increases} by 20\% as one goes from $3\times 2$-pt to $6\times 2$-pt data; this is because CMB lensing (present in $6\times 2$-pt) brings the matter density parameter down, and thus eases the mild tension between the higher value preferred by the $3\times 2$-pt data and the lower value preferred by DESI that led to a correspondingly tight error bar on $\Om$. \cref{fig:Om_s8_S8_LCDM} further illustrates these results, and compares the DESI+DESY3 ($6\times 2$-pt)  constraints to those obtained from the CMB temperature and polarisation (without CMB lensing\footnote{Recall that when we combine DESY3 ($6\times 2$-pt) data with the CMB, we use CMB without its lensing reconstruction (hence ``CMB-nl") in order not to double-count information with the CMB lensing present in the $6\times 2$-pt analysis; see \cref{sec:ext}.}).

It is also interesting to compare the combination of DESI and DESY3  ($6\times 2$-pt) measurements in \cref{eq:DESI_DESY3_LCDM} to those from DESY3  ($6\times 2$-pt) alone \cite{DES:2022urg}. The addition of DESI full shape and BAO data improves the constraints from DESY3  ($6\times 2$-pt) alone by about a factor of two in $\Om$ and $\sigma_8$, though only by about 10\% in $S_8$, the parameter to which gravitational-lensing surveys are sensitive.

We also investigate adding CMB (with no lensing reconstruction) data to the combination of DESI full-shape and BAO along with DESY3 ($6\times 2$-pt). We find 
%%%%%%%%%%%%%%%%%%%%%%%%%%%%%%%%%%%%%%%
\threeonesig[6cm]
{\Om &= 0.3009\pm 0.0034,\\[-0.45cm]}
{\sigma_8 &= 0.8028^{+0.0050}_{-0.0045},\\[-0.45cm]}
{H_0 &= (68.40\pm 0.27)\,\kmsMpc.}
{DESI (FS+BAO) + CMB-nl +  DESY3 ($6\times 2$-pt). \label{eq:DESI_CMB_DESY3_LCDM}}
%%%%%%%%%%%%%%%%%%%%%%%%%%%%%%%%%%%%%%%
We see that the addition of CMB-nl to the combination of DESI and DESY3 ($6\times 2$-pt) leads to about a factor of three improvement in the measurement error of cosmological parameters. Specifically, the matter density is determined to 1\% accuracy, $\sigma_8$ and $S_8$ are determined to 0.6\%, and the Hubble constant is pinned down to 0.4\%. These are the strongest constraints in the \lcdm\ model presented in this paper, and they show the remarkable power of modern survey data to measure key parameters of the cosmological model. We also find that the addition of DESI data improves the measurement of the Hubble constant by $\sim$20\% compared to CMB-nl+DESY3 ($6\times 2$-pt) data alone. Our constraints are further illustrated in the $\Om$--$\sigma_8$ plane in \cref{fig:Om_s8_LCDM}.
    
Our baseline measurement of the Hubble constant in \cref{eq:DESI_LCDM} indicates that DESI alone (helped with the BBN and $\nsten$ priors) prefers lower values of $H_0$, in agreement with independent measurements by the CMB \cite{Planck-2018-cosmology}. When DESI and the CMB (no lensing) are combined together and further helped with the DESY3 ($6\times 2$-pt) data, the central value of $H_0$ does not appreciably change but the errors decrease by about a factor of three relative to DESI alone; the resulting measurement of the Hubble constant (\cref{eq:DESI_CMB_DESY3_LCDM}) is in a $4.5\sigma$ tension with the much higher value preferred by the distance-ladder measurements that use Cepheid variables and nearby SN~Ia \cite{Riess:2021jrx}. We will study cosmological-parameter tensions in more detail in a dedicated supporting paper \cite{KP7s2-Tension}.

Note finally that the Hubble constant constraints determined in the FS+BAO analysis, and its combinations with external probes, depend on sound horizon physics. It will be of interest to compare our results to analyses that use sound-horizon-independent methods, and either marginalise over the sound horizon 
\cite{Farren:2021grl,DESI:BAO-free}, use only energy densities \cite{Krolewski:2024jwj}, or use power spectrum or correlation function features that depend on the epoch of matter-radiation equality \cite{Brieden:2022heh,Bahr-Kalus:2023ebd,DESI:TurnOver}. 

\subsection{$w_0w_a$CDM model}
\label{sec:DE}

The combination of DESI DR1 baryon acoustic oscilllations with cosmic microwave background and type Ia supernova datasets demonstrated a preference for a time-varying dark energy equation of state \cite{DESI2024.VI.KP7A}.  Here we report how these $w_0$--$w_a$ results are updated once we add full-shape information to the DESI BAO data. Moreover, we investigate whether generalising the cosmological model from \lcdm\ to \wowacdm\ loosens the constraints on the other parameters of interest, such as $\sigma_8$ and (as we will study in \cref{sec:neutrinos}) the sum of the neutrino masses.  

We study the time-varying dark energy equation of state in the parameterisation \cite{Chevallier:2001,Linder2003}
\begin{equation}
\label{eq:DE_EoS}
    w(a) =w_0+w_a(1-a),
\end{equation}
where $w_0$ and $w_a$ are the two beyond-\lcdm\ parameters describing the temporal evolution of the dark-energy equation of state, and $a$ is the scale factor. We make use of the parametrised post-Friedmann approach \cite{Fang:2008sn} to compute the dark energy perturbations when
calculating the CMB angular power spectrum.
We do not show the measurements derived from DESI (FS+BAO) data alone, as they are significantly affected by parameter projection effects (see \cite{DESI2024.V.KP5} and  \cref{sec:projection}). The reason that the projection effects are much more pronounced in the FS+BAO constraints in \wowacdm\ than in the equivalent BAO-alone analysis \cite{DESI2024.VI.KP7A} is the presence of many additional nuisance parameters in the full-shape analysis which allow additional freedom and open new degeneracy directions. Therefore, we only consider DESI full-shape clustering and BAO in combination with other data when testing the \wowacdm\ model.  We find that the combination of DESI and CMB is also subject to strong projection effects, but that further addition of data from type Ia supernovae removes them (see again \cref{sec:projection}).

\begin{figure*}
\includegraphics[width=\linewidth]{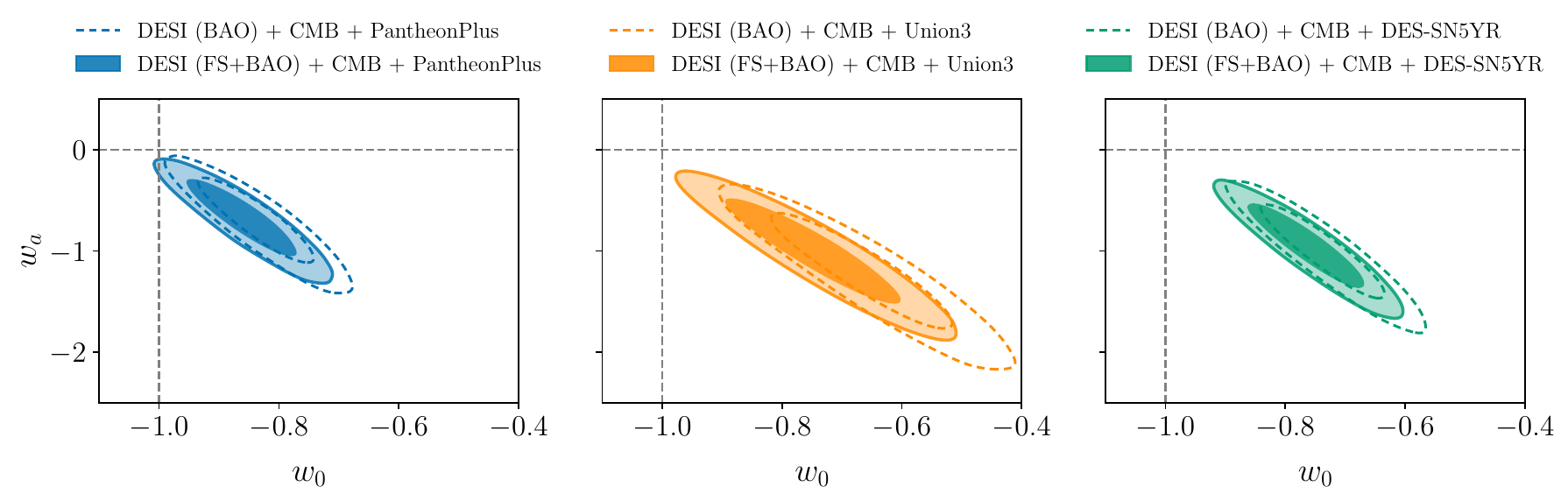} 
\caption{Constraints on \( w_0 \) and \( w_a \), assuming a \wowacdm\ model with a time-varying dark energy equation of state parameterisation (\cref{eq:DE_EoS}). The contours represent the 68\% and 95\% credible intervals. The solid blue, orange, and green contours represent the combination of DESI (FS+BAO) and CMB with three respective SN~Ia data sets: PantheonPlus, Union3 and DES-SN5YR. The dashed blue, orange, and green  contours show the same respective combinations, but with the DESI full-shape clustering and BAO replaced by DESI (BAO). The figure shows how the addition of the full-shape information to the BAO-only data improves the precision of the constraints. The measurements of these two parameters remain mutually consistent, and  prefer $ w_0 > -1 $ and $w_a < 0$.
}
\label{Fig:w0waCDM_w0-wa_constraints}
\end{figure*}

Therefore, when allowing the extra freedom in the expansion history allowed by the \wowacdm\ model, we only present results for the combination of our DESI data with the CMB as well as the various type Ia supernova datasets. With PantheonPlus, we find 
%%%%%%%%%%%%%%%%%%%%%%%%%%%
\twoonesig[5cm]
{w_0 &= -0.858\pm 0.061,}
{w_a &= -0.68^{+0.27}_{-0.23},}
{DESI (FS+BAO)+CMB\dataplus PantheonPlus. \label{eq:DESI_CMB_Pantheon_w0waCDM}}
%%%%%%%%%%%%%%%%%%%%%%%%%%%
The combination with Union3 supernova data results in
%%%%%%%%%%%%%%%%%%%%%%%%%%%
\twoonesig[4.5cm]
{w_0 &= -0.742\pm 0.096, }
{w_a &= -1.02^{+0.36}_{-0.32},}
{DESI (FS+BAO)+CMB\dataplus Union3. \label{eq:DESI_CMB_Union3_w0waCDM}}
%%%%%%%%%%%%%%%%%%%%%%%%%%%
Finally, the combination with DES-SN5YR gives
%%%%%%%%%%%%%%%%%%%%%%%%%%%
\twoonesig[4.5cm]
{w_0 &= -0.761\pm 0.065, }
{w_a &= -0.96^{+0.30}_{-0.26},}
{DESI (FS+BAO)+CMB\dataplus DES-SN5YR. \label{eq:DESI_CMB_DESY5_w0waCDM}}
%%%%%%%%%%%%%%%%%%%%%%%%%%%
These results are summarised in \cref{Fig:w0waCDM_w0-wa_constraints}, which shows that the outcomes of the combined analyses that include DESI with BAO-only information are consistent with those that include both the full-shape and the BAO information. Moreover, when the DESI (BAO)+CMB+SN~Ia combination is supplemented with the full-shape information from DESI, the constraints on the dark energy equation of state parameters improve: the dark energy Figure of Merit \cite{Albrecht:2006um}\footnote{The Dark Energy Task Force Figure of Merit (FoM) is defined as the inverse area of the 95\% posterior contour in the $w_0$--$w_a$ plane. For a Gaussian posterior, ${\rm FoM}\propto |\det \mathbf{C}|^{-1/2}$, where $\mathbf{C}$ is the projected $2\times 2$ covariance in the ($w_0, w_a$) subspace, and this is the definition we use to calculate the ratios of FoMs.} for combinations involving PantheonPlus, Union3, and DES-SN5YR increases by a factor of 1.16, 1.23 and 1.15 respectively. Thus, the respective credible-region areas in the $w_0$--$w_a$ plane are reduced by about 20\% when the full-shape data is added to the BAO.

\begin{figure*}
\centering
\includegraphics[width=10.0cm]{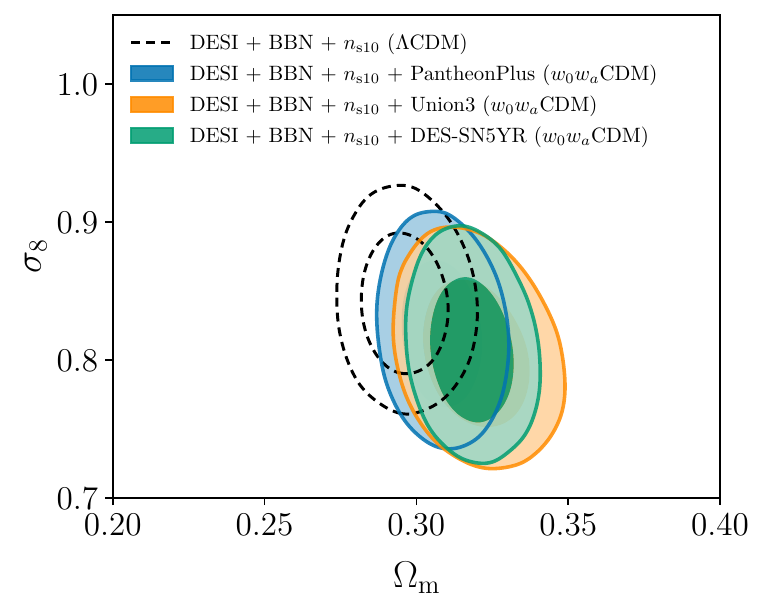} 
\caption{Constraints in the $\Om$--$\sigma_8$ plane illustrating the dependence on the assumed cosmological background. The dashed black contour shows the result of analysing DESI full-shape clustering and BAO in \lcdm\ model. The blue, orange and green contours represents the respective combinations of DESI with PantheonPlus, Union3 and DES Year-5 supernova datasets, all assuming the \wowacdm\ model. In all cases, we include the usual BBN and $\nsten$ priors. The colored contours highlight a modest shift to smaller $\sigma_8$ values as the cosmological background changes from \lcdm\ to \wowacdm. Overall, however, the constraints on both parameters remain consistent between all cases. 
}
\label{Fig:w0waCDM_sigma8-Om_constraints}
\end{figure*}

\begin{figure*}
\includegraphics[width=15.0cm]{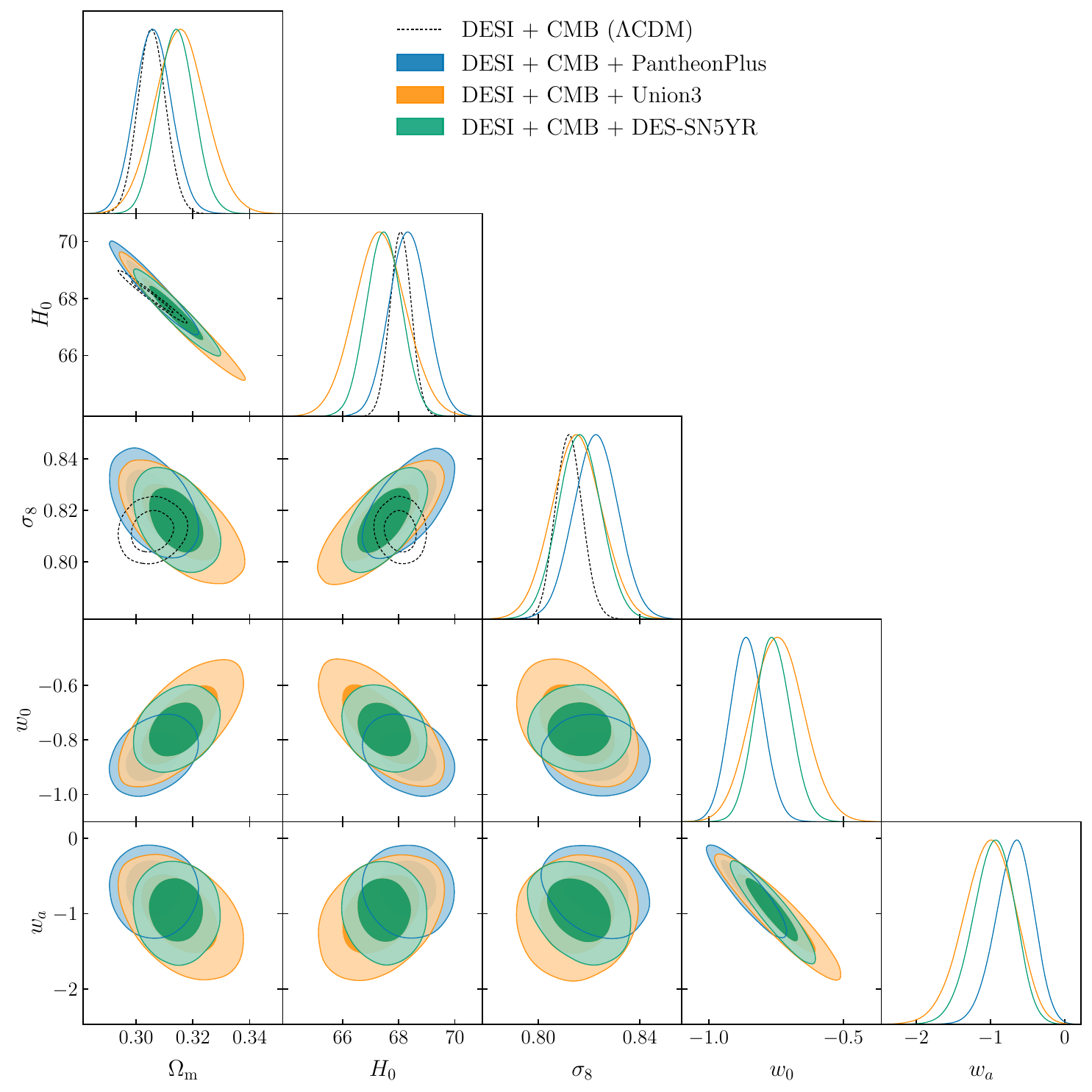}
\caption{Constraints in the \wowacdm\ model, projected onto the five-dimensional subspace spanned by $\Om, H_0, \sigma_8, w_0$ and $w_a$. We show the combination of DESI full-shape and BAO and the CMB with each of the three SN~Ia  datasets: PantheonPlus (blue), Union3 (orange) and DES-SN5YR (green). For comparison, the black dotted contours illustrate the constraints obtained from DESI and the CMB within \wowacdm.}
\label{Fig:w0waCDM_w0_wa_triangle_plot}
\end{figure*}

\cref{Fig:w0waCDM_w0-wa_constraints} also illustrates that, when combining our full-shape and BAO results with that from the cosmic microwave background and type Ia supernovae, the results remain fully consistent with the same probe combination that contains DESI BAO-only \cite{DESI2024.VI.KP7A}, and continue to indicate a preference for a departure from the \lcdm\ values of $(w_0=-1, w_a=0)$. Calculating the difference between the maximum \textit{a posteriori} value of the \wowacdm\ models in our chains, and the MAP value of the models that enforce \lcdm\ $(w_0=-1, w_a = 0)$, we find values of $\Delta \chi_\mathrm{MAP}^2=-8.8$, $-14.5$ and $-17.5$ for the combinations of DESI and CMB with PantheonPlus, Union3 and DES-SN5YR respectively. These values of $\Delta \chi_\mathrm{MAP}^2$ correspond to preferences for  \wowacdm\ over \lcdm\ at the significance levels of $2.5\sigma$ (PantheonPlus), $3.4\sigma$ (Union3), and $3.8\sigma$ (DES-SN5YR). These preferences are similar to the ones we found for the combination of DESI BAO-only data with CMB and SN~Ia \cite{DESI2024.VI.KP7A}. Note also that the change in the preference for departures from \lcdm\ in the combination of DESI and CMB with Union3 data, when going from BAO to full-shape plus BAO, is smaller than what \cref{Fig:w0waCDM_w0-wa_constraints} visually implies because of the noticeable projection effects in this combination (compare the mean and MAP values of $w_0$ and $w_a$ for this combination of probes in \cref{tab:parameter_table1}).

\cref{Fig:w0waCDM_sigma8-Om_constraints} illustrates the dependence of the measurements of $\Om$ and $\sigma_8$ on the cosmological background assumed. We show the DESI full-shape plus BAO constraints in the $\Om$--$\sigma_8$ plane assuming \lcdm\ background, and also constraints in the $w_0 w_a$CDM model for DESI in combination with each of the three SN~Ia samples. Overall, the measurements of $\Om$ and $\sigma_8$ are consistent for all four cases shown. In more detail, the contours for the $w_0w_a$CDM model shift slightly toward higher $\Om$ and lower $\sigma_8$ values compared to the DESI-only $\Lambda$CDM results presented in \cref{eq:DESI_LCDM}. The numerical results for these measurements are presented in \cref{tab:parameter_table1}. Note in particular that the values of $S_8$ in the $w_0 w_a$CDM background from the combination of DESI and SN~Ia datasets remain fully consistent with its value in $\Lambda$CDM from DESI alone, as the changes in $\sigma_8$ and $\Om$ (that enter in the definition of $S_8$) effectively cancel out. 

\cref{Fig:w0waCDM_w0_wa_triangle_plot} shows a more detailed scan of the parameter space in our \wowacdm\ analysis, showing the projection onto the five-dimensional subspace spanned by $\Om, H_0, \sigma_8, w_0$ and $w_a$. We show results for the combination of DESI (FS+BAO) and CMB with each of the three SN~Ia datasets. The numerical constraints are presented in \cref{tab:parameter_table1}. When the cosmological background changes from \lcdm\ to \wowacdm, the central values in $\Om$, $\sigma_8$/$S_8$ and $H_0$ remain unchanged within the errors. The error bars in these parameters increase but remain small, with percent-level precision in each for all three DESI+CMB+SN~Ia combinations. This shows the robustness of our constraints on these key parameters to variations in the underlying cosmological model.

In summary, DESI full-shape clustering and BAO data, in combination with the CMB and SN~Ia, continue to show hints of a departure from the \lcdm\ model. 
% \dragan{new sentence:} 
The degree of preference for this departure from the standard cosmological model depends on the choice of the SN~Ia dataset. 
This preference has
%, however, not been seen 
already been investigated in the recent full-shape re-analysis of BOSS data  that combines it with similar external data as our study \cite{Chen:2024vuf}, and it will be interesting to investigate in detail how the choice of dataset and full-shape methodology affects the results. Studying models with more freedom in the dark-energy sector than allowed in \wowacdm\ is also promising \cite{DESI:2024aqx,DESI:2024kob}, especially as the data get better. Looking ahead, DESI's forthcoming Year-3 data analysis will shed significant new light on dynamical dark energy.

\section{Neutrino constraints}
\label{sec:neutrinos}

In this section we exploit the DESI full-shape data combined with other datasets in order to set limits on the sum of neutrino masses and on the number of relativistic species in the early universe. 

\subsection{Sum of neutrino masses}

The existence of massive neutrinos, implied by the discovery of neutrino oscillations \cite{Fukuda98,Ahmad02,Eguchi03}, is direct evidence for physics beyond the Standard Model. Major efforts are underway to constrain neutrino properties in laboratory experiments, but neither the ordering of the neutrino masses nor their absolute scale is known. Oscillation experiments are sensitive to the mass squared differences \cite{Esteban20,Nufit24,Capozzi21,DeSalas21} and set a lower bound on the sum of the three neutrino masses. If the smallest mass splitting is between the lowest mass eigenstates, neutrino masses are said to have the normal ordering (NO) and satisfy $\sum m_\nu\geq\SI{0.059}{\eV}$. The other possibility, known as the inverted ordering (IO), implies $\sum m_\nu\geq\SI{0.10}{\eV}$. The strongest model-independent upper bound on the absolute mass scale comes from the KATRIN experiment \cite{Katrin:2024tvg}, which constrains the effective electron anti-neutrino mass to $m_\beta<\SI{0.45}{\eV}$ (90\% CL). Assuming three quasi-degenerate neutrinos, this is equivalent to $\sumnu<\SI{1.35}{\eV}$ (90\% CL), which is an order of magnitude higher than typical upper limits from cosmology.

Cosmological probes are sensitive to a number of distinctive signatures of cosmic background neutrinos, which enable independent and complementary constraints on $\sum m_\nu$ \cite{Lesgourgues:2006,Wong:2011,Abazajian:2016} and can be broadly separated into effects on the background expansion and on the growth of fluctuations. At the background level, massive neutrinos affect the expansion history in a unique way, contributing as radiation in the early universe and as non-relativistic matter at recent epochs. Neutrino masses can be tightly constrained from this signature alone. Massive neutrinos also have a strong effect on the growth of cosmic structure. After becoming non-relativistic, neutrinos retain large thermal velocities and cannot be contained in regions smaller than a typical free-streaming length, which is  inversely proportional to their mass. Since, at this stage, neutrinos contribute fully to the expansion as non-relativistic matter, but only partially to the clustering, the growth of density perturbations is reduced on small scales. This is manifested as a scale-dependent suppression of the matter power spectrum, which scales as $\Delta P(k)/P(k)\propto -\Omega_\nu/\Om$ \cite{Hu:1998,Lesgourgues:2006,Kiakotou:2008} and affects equally the broadband shape of the power spectrum and the amplitude of the BAO, with constraints from the latter effect being potentially less prone to parameter projection effects \cite{Noriega:2024lzo}.

The DESI full-shape power spectrum analysis allows the sum of neutrino masses to be constrained independently of the CMB. Assuming a $\Lambda$CDM background and three degenerate neutrino species, we find an  upper bound

%%%%%%%%%%%%%%%%%%%%%%%%%%%%%%
\onetwosig[7cm]{\sumnu < 0.409\eV}{{DESI (FS+BAO)+BBN+$\nsten$}}{. \label{eq:mnu_DESI}} 
%%%%%%%%%%%%%%%%%%%%%%%%%%%%%%

\noindent
We reiterate that we adopt two external priors when DESI is not combined with CMB data: an external BBN prior on $\Omega_\text{b}h^2$ and a weak prior on the spectral index, $\ns$, that corresponds to ten times the uncertainty $(10\sigma)$ from \emph{Planck}. On the scales measured by DESI, the broadband suppression from neutrinos is degenerate with $\ns$. Moreover, $\ns$ is also degenerate with $H_0$ and with $\Omega_\text{m}$. Adding a stronger $(1\sigma)$ Gaussian prior on $\ns$, representing a limited use of CMB information, improves the upper bound to

%%%%%%%%%%%%%%%%%%%%%%%%%%%%%%
\onetwosig[8cm]{\sumnu < 0.300\eV}{{DESI (FS+BAO)+BBN+tight $\ns$ prior}}{. \label{eq:mnu_DESI_ns}} 
%%%%%%%%%%%%%%%%%%%%%%%%%%%%%%

\begin{figure*}
\begin{center}
\includegraphics[height=13cm]{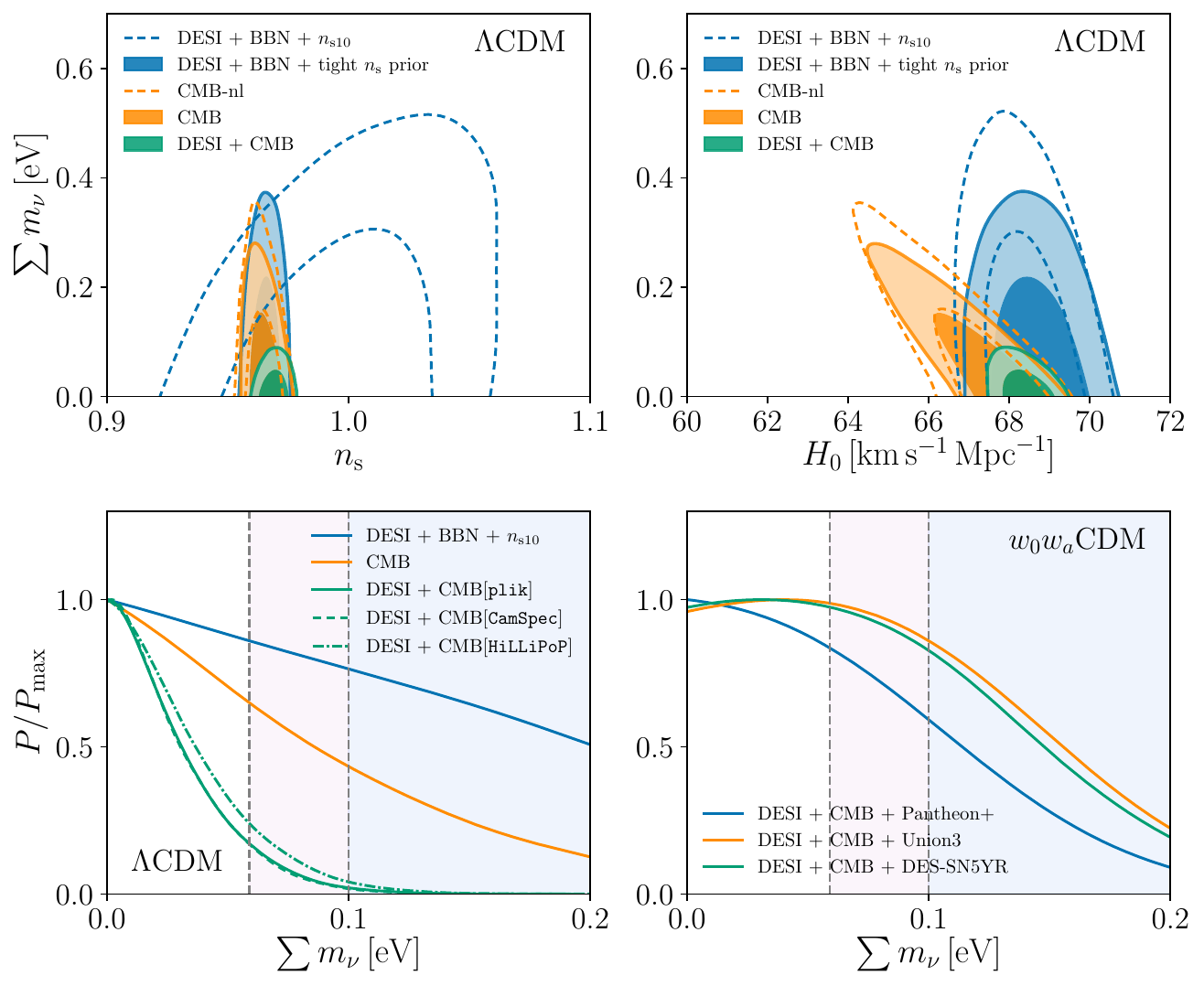}
\end{center}
\caption{
\emph{Top left panel}: constraints in the $\ns$-$\sumnu$ plane. The blue dashed contours show the 68\% and 95\% credible intervals for the fiducial DESI (FS+BAO) dataset, accompanied, as usual, by the BBN prior on $\Ob h^2$ and a loose prior on $\ns$. The filled blue contours illustrate the improvement when the $\ns$ prior is tightened to be that from \emph{Planck} (rather than 10 times weaker). The dashed orange contours show the results from CMB without the lensing reconstruction, while the filled orange contours show the constraints from CMB with lensing. Finally, the green contours show the DESI+CMB combination. \emph{Top right panel}: constraints in the $H_0$-$\sumnu$ plane for the same data combinations as in the top left panel, illustrating that the DESI+CMB combination breaks the geometric degeneracy between $H_0$ and $\sumnu$. \emph{Bottom left panel}: one-dimensional posteriors on the sum of the neutrino masses. We show constraints from DESI (FS+BAO) alone, CMB alone, and DESI+CMB for three alternative choices of the CMB likelihood. The minimal masses for the normal or inverted mass ordering scenarios, corresponding respectively to  $\sumnu\geq0.059\eV$ and $\sumnu\geq0.10\eV$, are shown by the vertical dashed lines and the shaded regions. \emph{Bottom right panel}: same as for the bottom left panel, but for the \wowacdm\ background and showing constraints from the combination of DESI, CMB, and SN Ia as labelled.
}
\label{fig:neutrino_figure}
\end{figure*}

\noindent
This is similar to the constraint from the CMB without CMB lensing, $\sum m_\nu<\SI{0.265}{\eV}$ (95\%), as can be seen from the top left panel of \cref{fig:neutrino_figure}, but relies on the growth of fluctuations instead of primarily on effects on the background expansion (see \cite{Loverde:2024} for an analysis of the contributions of geometric and growth information from the CMB). The physics behind these constraints will be addressed in greater detail in \cite{DESI2024.ValueAddedNu}. The combination of CMB temperature and polarisation data and CMB lensing, as before simply denoted as ``CMB'', yields an upper bound of $\sum m_\nu<\SI{0.218}{\eV}$ (95\%). The tightest limits are obtained from the full combination of DESI and CMB measurements, demonstrating their strong complementarity. First, using CMB and DESI BAO only yields

%%%%%%%%%%%%%%%%%%%%%%%%%%%%%%
\onetwosig[6cm]{\sumnu < 0.082\eV}{{DESI (BAO only)+CMB}}{. \label{eq:mnu_DESI_BAO_CMB}} 
%%%%%%%%%%%%%%%%%%%%%%%%%%%%%%

\noindent
Compared to the equivalent result for the same data combination reported in \cite{DESI2024.VI.KP7A}, this figure is slightly higher due to a change in the external CMB lensing likelihood as we switched from version \texttt{v1.1} of the ACT lensing likelihood to version \texttt{v1.2}. Nevertheless, it represents the strongest cosmological bound from CMB and BAO information only. Adding the full-shape information improves the upper bound further to

%%%%%%%%%%%%%%%%%%%%%%%%%%%%%%
\onetwosig[6cm]{\sumnu < 0.071\eV}{{DESI (FS+BAO)+CMB}}{. \label{eq:mnu_DESI_CMB}} 
%%%%%%%%%%%%%%%%%%%%%%%%%%%%%%

\noindent
The tight limit in \cref{eq:mnu_DESI_BAO_CMB} from DESI BAO + CMB arises from the preference of DESI data for high values of $H_0$ and low values of $\Om$, which suppresses the bounds on $\sum m_\nu$ due to the geometric degeneracy between these parameters. This trend is further reinforced by the DESI full-shape analysis, which improves the precision in $H_0$ and especially $\Om$ once combined with DESI BAO. Hence, the improvement in the constraint seen in \cref{eq:mnu_DESI_CMB} is not directly associated with the suppression of the power spectrum, but with a greater pull towards low $\Om$ and high $H_0$. This is illustrated in the top right panel of \cref{fig:neutrino_figure}, which shows how the degeneracy in the $H_0$-$\sumnu$ plane is broken by the DESI + CMB combination.

\begin{table}
    \centering
    \small
    \begin{tabular}{lcccc}
    \toprule
    \midrule
    model / dataset & $\Omega_{\mathrm{m}}$ & $H_0$ [$\kmsMpc$] & $\sum m_\nu\,[\mathrm{eV}]$ & $N_{\mathrm{eff}}$\\
    \midrule
    %%%%%%
    $\mathbf{\Lambda}${\bf CDM+}$\boldsymbol{\sumnu}$ &&&&\\
    DESI (FS+BAO)+BBN+$\nsten$ &  $0.2991^{+0.0098}_{-0.011}$    &  $68.40\pm 0.78$    &  $< 0.409$   &---\\  
    DESI+CMB &  $0.3026\pm 0.0052$    &  $68.35\pm 0.41$    &  $< 0.071$   &---\\  
    \midrule
    %%%%%%
    $\mathbf{\Lambda}${\bf CDM+}$\boldsymbol{\Neff}$ &&&&\\
    DESI+CMB &  $0.3028\pm 0.0059$        &  $68.9\pm 1.1$    &---  & $3.18\pm 0.16$ \\
    \midrule
    %%%%%%%
    $\boldsymbol{w_0w_a}${\bf CDM+}$\boldsymbol{\sumnu}$ &&&&\\
    % DESI+CMB &           &           &           \\
    DESI+CMB+PantheonPlus &   $0.3064\pm 0.0067$        &    $68.33\pm 0.68$       &  $< 0.175$         &---\\
    DESI+CMB+Union3 &   $0.3167\pm 0.0095$        &    $67.30\pm 0.93$       &  $< 0.201$         &---\\
    DESI+CMB+DES-Y5SN &   $0.3151\pm 0.0067$        &    $67.45\pm 0.63$       &  $< 0.196$         &---\\
    \midrule
    %%%%%%%
    $\boldsymbol{w_0w_a}${\bf CDM+}$\boldsymbol{\Neff}$ &&&&\\
    % DESI+CMB &          &           &---&           \\
    DESI+CMB+PantheonPlus &    $0.3068\pm 0.0066$       &   $68.0\pm 1.1$        &---&     $2.97\pm 0.17$      \\
    DESI+CMB+Union3 &    $0.3167\pm 0.0093$       &   $66.8\pm 1.2$        &---&     $2.94\pm 0.17$      \\
    DESI+CMB+DES-Y5SN &    $0.3152\pm 0.0065$       &   $67.0\pm 1.0$        &---&     $2.94\pm 0.17$      \\
    \midrule
    \bottomrule
    \end{tabular}
    \caption{
    Cosmological parameter estimates and constraints from DESI DR1 full-shape clustering and BAO data, in combination with external datasets, when considering extensions in the neutrino sector of the \lcdm\ and $w_0w_a$CDM models (``DESI'' in the table stands for DESI DR1 (FS + BAO)). Results with two-sided error bars refer to the marginalised means and 68\% credible intervals; upper bounds on $\sumnu$ refer to  95\% limits. All constraints on $\sumnu$ assume a model with three degenerate mass eigenstates and a minimal prior, $\sumnu>0$ eV. The empty $\sumnu$ and $\Neff$ fields indicate that fixed values of $\sumnu=0.06\eV$ and $\Neff=3.044$ respectively were adopted.
    \label{tab:parameter_table2}}
\end{table}

The marginalised posterior distribution of $\sum m_\nu$ is shown in the bottom left panel of \cref{fig:neutrino_figure}. As was the case for the DESI BAO analysis \cite{DESI2024.VI.KP7A}, the posterior peaks at a value near $\sum m_\nu=\SI{0}{\eV}$, which is excluded by neutrino oscillations. This is true for all data combinations reported above, including the CMB-independent result in \cref{eq:mnu_DESI}. Although similar behavior had already been seen in \emph{Planck} and SDSS data \cite[e.g.][]{Planck14,BOSS:2016wmc,eBOSS:2020yzd,Noriega:2024lzo}, the results have always been compatible with the oscillation constraints. 

Since the release of the DESI BAO results, a number of recent studies have identified stronger tensions between the constraints from cosmology and neutrino oscillations, either by combining the DESI BAO measurements with additional data sets \cite{Jiang24,Wang24} or by considering the possibility of neutrino masses beyond the experimental limits, including apparent negative values, as an indicator of systematics, new neutrino properties or a non-standard cosmological expansion \cite{Craig24,Elbers24,Green24}. Compared to the DESI (BAO) analysis, the full-shape information leads to a $15\%$ stronger constraint and a slight increase in the tension. Nevertheless, our baseline DESI + CMB result remains compatible with the lower bound for the normal ordering at the  $\sim$2$\sigma$ level. Results from the full-shape analysis for extended neutrino models will be presented in \cite{DESI2024.ValueAddedNu}.

The bounds given above were all obtained under the assumption of three degenerate neutrino species with a prior that $\sum m_\nu>0$. The $95\%$ upper limits would increase considerably if more restrictive priors, motivated by neutrino oscillations ($\sum m_\nu\geq\SI{0.059}{\eV}$ or $\sum m_\nu\geq\SI{0.10}{\eV}$ depending on the ordering), were imposed \cite{DESI2024.VI.KP7A}. This is a consequence of the fact that much of the posterior volume is in the unphysical range ($\sum m_\nu<\SI{0.059}{\eV}$). The preference for the normal ordering over the inverted ordering from DESI + CMB stands at a modest $\Delta\chi^2_\mathrm{MAP}\simeq -2$ level. The implications for the neutrino mass ordering of a combined analysis of cosmological and laboratory data will be discussed in detail in a forthcoming publication \cite{DESI2024.ValueAddedNu}.

A potential systematic affecting neutrino mass bounds is the so-called CMB lensing anomaly \cite{Allali24,Choudhury24,Craig24,Elbers24,NaredoTuero24}, which refers to a small oscillatory feature in the \emph{Planck} data that could be explained by additional gravitational lensing of the CMB \cite{Calabrese08,Renzi18,Mokeddem23,Planck-2018-cosmology}. Two independent analyses of the latest \emph{Planck} PR4 data release, the high-$\ell$ \texttt{CamSpec} likelihood \cite{Efstathiou:2021,Rosenberg:2022} and the low-$\ell$ \texttt{LoLLiPoP} and high-$\ell$ \texttt{HiLLiPoP} likelihoods \cite{Tristram:2021,Tristram:2023}, are less affected by this anomaly. To investigate the robustness of our constraints, we repeat the analysis but replace the high-$\ell$ PR3 \texttt{plik} (TTTEEE) likelihood with PR4 \texttt{CamSpec}. This combination yields a bound that is nearly identical to our baseline result in \cref{eq:mnu_DESI_CMB},

%%%%%%%%%%%%%%%%%%%%%%%%%%%%%%
\onetwosig[8.5cm]{\sumnu < 0.069\eV}{{DESI (FS+BAO) + CMB[\texttt{CamSpec}]}}{. \label{eq:mnu_DESI_CMB_CamSpec}} 
%%%%%%%%%%%%%%%%%%%%%%%%%%%%%%

\noindent
Replacing both the low-$\ell$ \texttt{simall} (EE) likelihood with \texttt{LoLLiPoP} and the high-$\ell$ \texttt{plik} likelihood with \texttt{HiLLiPoP} has a somewhat larger effect, relaxing the bound to 

%%%%%%%%%%%%%%%%%%%%%%%%%%%%%%
\onetwosig[8.5cm]{\sumnu < 0.081\eV}{{DESI (FS+BAO) + CMB[\texttt{HiLLiPoP}]}}{. \label{eq:mnu_DESI_CMB_hillipop}} 
%%%%%%%%%%%%%%%%%%%%%%%%%%%%%%

\noindent
The associated posteriors are shown in the bottom left panel of \cref{fig:neutrino_figure}. Overall, the posteriors are fairly consistent, but the likelihoods that are least affected by the lensing anomaly (\texttt{LoLLiPoP} and \texttt{HiLLiPoP}) yield a slightly greater upper bound.

It is important to note that constraints on $\sum m_\nu$ depend strongly on the assumed dark energy model. Due to parameter degeneracies, allowing the dark energy energy equation of state, $w$, to vary can considerably relax the upper bound on $\sum m_\nu$ \cite{Hannestad:2005}, although this depends on the conditions imposed on $w$ \cite{Vagnozzi:2018,RoyChoudhury:2018vnm,Jiang24}. In this section, as in \cref{sec:cosmo_constraints}, we restrict attention to the $w_0w_a$CDM model in which the dark energy equation of state is a function of time described by two parameters. For the combination with DES-SN5YR in \wowacdm, the upper bound on the neutrino mass relaxes to

%%%%%%%%%%%%%%%%%%%%%%%%%%%%%%%%%%%%%%%%%%
\threeonesig[5cm]
{\sumnu&  < 0.196\eV \;\;(95\%),}
{w_0   & = -0.753\pm 0.070,}
{w_a   &= -1.02^{+0.37}_{-0.29},}
{DESI (FS+BAO) + CMB + DES-SN5YR, \label{eq:w0wa_mnu_DESI+ext}}
%%%%%%%%%%%%%%%%%%%%%%%%%%%%%%%%%%%%%%%%%%%

\noindent
while the constraints on $w_0$ and $w_a$ do not differ significantly from those obtained for fixed $\sum m_\nu$. The bottom right panel of \cref{fig:neutrino_figure} shows the marginalised posterior distribution of $\sumnu$ in the $w_0w_a$CDM model for DESI + CMB combined with the three supernova datasets: PantheonPlus, Union3, and DES-SN5YR. In the case of Union3 and DES-SN5YR, the maximum of the posterior is recovered near the physical mass range. However, in all cases, the data can accommodate larger neutrino masses, alleviating the tension with neutrino oscillations \cite{Elbers24,Choudhury:2024}. These results are in line with those of the DESI (BAO) analysis \cite{DESI2024.VI.KP7A}. The constraints for the combinations with Union3 and PantheonPlus are given in Table~\ref{tab:parameter_table2}.

\subsection{Number of relativistic species}

We also report constraints on the effective number of relativistic species, $\Neff$. This parameter is defined in terms of the energy density, $\rho_\nu$, due to neutrinos before their non-relativistic transition, which is given by
\begin{align}
    \rho_\nu = \Neff\, \frac{7}{8}\left(\frac{4}{11}\right)^{4/3}\rho_\gamma,
\end{align}

\noindent
where $\rho_\gamma$ is the energy density of photons. The standard model prediction for the non-instantaneous decoupling of three neutrino species is $\Neff=3.044$ \cite{Mangano:2005,deSalas:2016,Akita:2020,Froustey:2020,Bennett:2020}. However, the possibility of new degrees of freedom contributing additional dark radiation adding to $\Neff$  motivates extending the $\Lambda$CDM model to include $\Neff$ as a free parameter. Although $\Neff$ can be constrained through its imprint on the shape and phases of the BAO oscillations \cite{2017JCAP...11..007B,2019NatPh..15..465B,DESI2024.PhaseShift}, DESI primarily contributes constraining power by breaking the degeneracies with $H_0$ and $\Om$. For the $\Lambda$CDM+$\Neff$ model, we obtain a constraint from DESI BAO and CMB data of

%%%%%%%%%%%%%%%%%%%%%%%%%%%%%%
\oneonesig[5cm]{\Neff = 3.07\pm0.17}{{DESI (BAO)+CMB}}{. \label{eq:Neff_DESIBAO_CMB}} 
%%%%%%%%%%%%%%%%%%%%%%%%%%%%%%

\noindent
As was the case for the bounds on $\sum m_\nu$, this value differs slightly from the constraint for the same data combination reported in \cite{DESI2024.VI.KP7A} due to the switch from version \texttt{v1.1} of the ACT CMB lensing likelihood to version \texttt{v1.2}. The addition of the full-shape information leads to

%%%%%%%%%%%%%%%%%%%%%%%%%%%%%%
\oneonesig[5cm]{\Neff = 3.18\pm0.16}{{DESI (FS+BAO)+CMB}}{. \label{eq:Neff_DESI_CMB}} 
%%%%%%%%%%%%%%%%%%%%%%%%%%%%%%

\noindent
This amounts to a slight reduction in uncertainty and an upward shift in the central value relative to the constraint from DESI (BAO) + CMB  (\cref{eq:Neff_DESIBAO_CMB}), which can be attributed to the preference of DESI data for high $H_0$ and low $\Om$. 

Finally, we present the constraints on $\Neff$ in the $w_0w_a$CDM model in \cref{tab:parameter_table2}. Compared to $\Lambda$CDM, the uncertainty on $\Neff$ increases only slightly and the constraints remain within $1\sigma$ of $\Neff=3.044$.

%%%%%%%%%%%%%%%%%%%%%%%%%% MG SECTION %%%%%%%%%%%%%%
\section{Modified-gravity  
%$\mu_0$, $\Sigma_0$ 
constraints  }
\label{sec:MG}
%%%%%%%%%%%%%%%%%%%%%%%%%%%%%%%%%%%%%%%%%%%%%%%%%%%%%

DESI full-shape clustering data are sensitive to the growth of large-scale structure, and can hence constrain deviations from general relativity, which we analyse here. We briefly describe the formalism that we use for the modified-gravity parametrisation and then provide results from DESI alone and in combination with other available data sets.   

%%%%%%%%%%%%%%%%%%%%%%%%%%%
\subsection{Modified-gravity formalism and parameterisation}
\label{sec:MGparameterisation}
%%%%%%%%%%%%%%%%%%%%%%%%%%%%%

A common and promising approach to testing deviations from general relativity (GR) is to add physically motivated phenomenological parameters to the perturbed Einstein's gravitational field equations and test the deviations of such parameters from their GR predicted values. Whilst many such modified-gravity (MG) parameterisations have been proposed in the literature (see e.g. the reviews  \cite{Clifton:2011jh,Koyama:2015vza,Ishak:2018his} and references therein), we focus here on one that is based on the coupling of gravitational potentials to the source content of spacetime. 

In the conformal Newtonian gauge, the flat Friedmann-Lema\^{i}tre-Robertson-Walker metric with scalar perturbations can be written as
\begin{equation}
ds^2=a(\tau)^2[-(1+2\Psi)d\tau^2+(1-2\Phi)\delta_{ij}dx^i dx^j],
\label{eq:line-element}
\end{equation}
where $\Psi$ and $\Phi$ are two gravitational potentials,  $\tau$ is the conformal time, $x^i$ and $x^j$ are spatial coordinates, and the sum over $i$ and $j$ is implied. 

The Einstein field equations applied to the line element, \cref{eq:line-element}, yield two equations describing the coupling and evolution of the gravitational field potentials.
%that can be written in Fourier space. 
The first equation relates the potential $\Psi$ to the space-time sources and reduces at late times (i.e. in the absence of anisotropic stresses) to  
\begin{equation}
k^2\Psi = -4\pi G a^2 \mu(a,k) \sum_i\rho_i\Delta_i,
\label{eq:definition_mu}
\end{equation}
where $\rho_i$ is the density of matter species $i$, and $\Delta_i$ is its gauge-invariant, rest-frame overdensity whose evolution describes the growth of inhomogeneities. The phenomenological MG function $\mu(a,k)$ 
%\cite{Zhang:2007nk,Amendola:2007rr} 
is added to modify the strength of the gravitational interaction and thus the growth rate of structure. 

The second perturbed Einstein equation relates the two gravitational potentials, $\Psi$ and $\Phi$ and their coupling to source energy densities and stress shear. In the late-time universe and assuming general relativity, it is expected that the anisotropic stress becomes negligible and that the two potentials are nearly equal at the present time. However, this may not be the case in modified-gravity models, where the two potentials can be different and the equation takes the form 
$(\Phi- \eta(a,k)\Psi) \approx 0$ where the gravitational slip function $\eta(a,k)$ parametrizes a possible deviation from GR.  Combining this expression with \cref{eq:definition_mu}, one gets the equation that is particularly useful for the motion of massless particles in a gravitational field (and hence for, e.g., gravitational lensing) and reads  
\begin{equation}
k^2(\Phi+\Psi)=-8\pi G a^2 \Sigma (a,k) \sum_i\rho_i\Delta_i,
\label{eq:definition_Sigma}
\end{equation}
where we introduced the MG function $\Sigma(a,k) \equiv \mu(a,k)(\eta(a,k)+1)/2$. On the left-hand side, 
$(\Phi+\Psi)$ is equal to twice the so-called Weyl potential that governs
the motion of massless particles while, on the right-hand side, the function $\Sigma(a,k)$ modifies the equation from its general-relativistic form.  
In general relativity, $\Sigma(a, k)=\mu(a, k)=1$.

\begin{figure*}
\centering
\begin{tabular}{c c}
{\includegraphics[width=7.6cm]{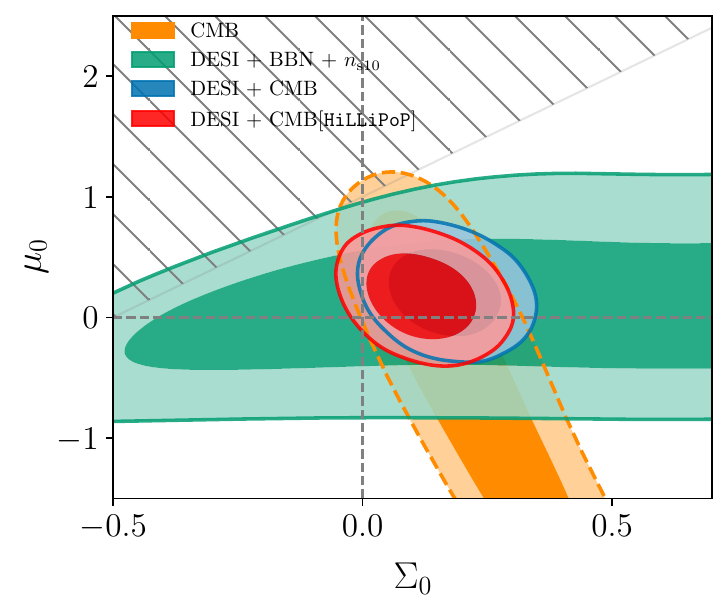}}
{\includegraphics[width=7.6cm]{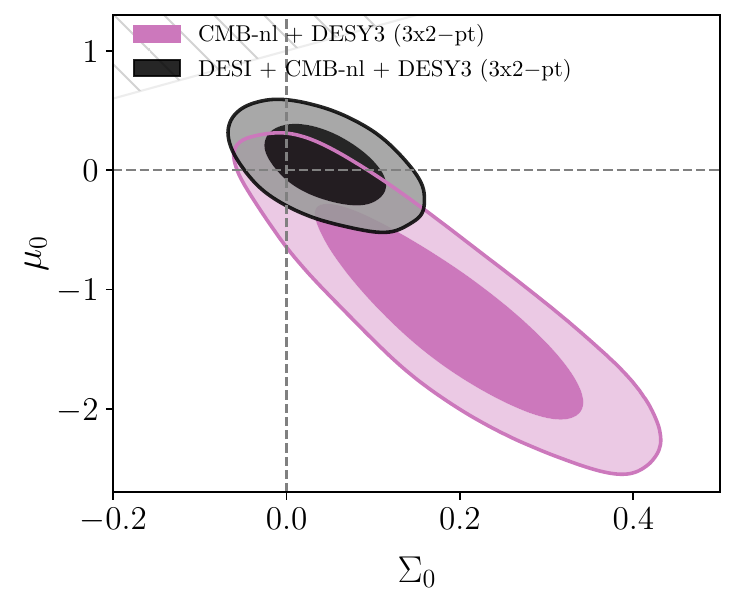}} 
\end{tabular}
\caption{
68\% and 95\% credible-interval constraints on modified-gravity parameters $\mu_0$ and $\Sigma_0$, assuming a \lcdm\ background. 
\emph{Left panel}: We show the constraints from the CMB alone in orange, and those from DESI full-shape clustering and BAO in green. The blue contour shows the constraints from the combination of DESI and the CMB with our fiducial \texttt{plik} likelihood, while the red contour shows the same with the alternate \texttt{LoLLiPoP}-\texttt{HiLLiPoP} CMB likelihood. Note that DESI data constraints are consistent with, and centered around, the general relativistic value of $\mu_0=0$, and this consistency is maintained once external data is added. Whilst DESI alone does not directly constrain $\Sigma_0$, when it is added to other datasets, like the CMB, it breaks other parameter degeneracies and helps tighten the constraints on $\Sigma_0$ (see full discussion of results in \cref{sec:MG_constraints}). {\emph{Right panel}: The purple contour shows the combination of CMB-nl with the galaxy clustering and weak lensing ($3\times 2$-pt) likelihood from DESY3, whilst the black contour shows the same combination with the addition of DESI.}
In both panels, the shaded area on the top left shows the hard prior $\mu_0 < 2 \Sigma_0 + 1$ that is imposed due to computational limitations of publicly-available modified-gravity codes based on \texttt{CAMB} (any overlap of the contours with the region excluded by the prior is an artefact of the KDE smoothing in the \texttt{getdist} plotting software); this prior does not affect our main results from the combination of probes (see \cref{sec:MG_constraints}).
} 
\label{Fig:MG_constraints}
\end{figure*}

%%%%%%% BEGIN MG TABLE %%%%%%%%%%%%%%%%%%%
\begin{table}
\centering
\resizebox{\columnwidth}{!}{%
    \small 
\setcellgapes{3pt}\makegapedcells  % for makecell command below     
\renewcommand{\arraystretch}{2.1} % row spacing
    \begin{tabular}{lccccc}
    \toprule
    \midrule
    \multirow{2}{*}{model/dataset} & \multirow{2}{*}{$\Om$} &  \multirow{2}{*}{$\sigma_8$} & $H_0$ & \multirow{2}{*}{$\mu_0$} & \multirow{2}{*}{$\Sigma_0$}\\[-0.3cm]
     & & & [${\rm km/s/Mpc}$] & & \\
    \midrule\midrule
    {\bf Flat} $\boldsymbol{\mu_0\Sigma_0}${\bf $\Lambda$CDM} &&&&\\[-0.2cm] 
     DESI (FS+BAO)+ BBN+$\nsten$

& $0.2957\pm 0.0097$ & $0.839\pm 0.034$ & $68.53\pm 0.75$ & $0.11^{+0.45}_{-0.54}$  & no constraint\\
     
    \hdashline    
    CMB-nl              
& $0.3041\pm 0.0093$ & $0.742^{+0.13}_{-0.092}$ & $68.21\pm 0.71$ & $-0.66^{+1.5}_{-0.83}$ & $0.47^{+0.16}_{-0.22}$ \\

  CMB-nl [\texttt{HiLLiPoP}]   & $0.3060\pm 0.0076$ & $0.737^{+0.13}_{-0.084}$ & $67.93\pm 0.57$ & $-0.73^{+1.4}_{-0.79}$ & $0.23^{+0.13}_{-0.20}$ \\

    \hdashline    

    DESI+CMB-nl    & $0.2985\pm 0.0055$ & $0.822\pm 0.024$ & $68.63\pm 0.43$ & $0.23\pm 0.24$ & $0.388^{+0.11}_{-0.086}$ \\

    DESI+CMB       & $0.3023\pm 0.0053$ & $0.824\pm 0.024$ & $68.32\pm 0.41$ & $0.21\pm 0.24$ & $0.166\pm 0.074$ \\

    DESI+CMB-nl [\texttt{HiLLiPoP}]    & $0.3006\pm 0.0051$ & $0.824\pm 0.024$ & $68.33\pm 0.40$ & $0.22\pm 0.24$ & $0.148^{+0.097}_{-0.12}$ \\

    DESI+CMB [\texttt{HiLLiPoP}]     & $0.3028\pm 0.0050$ & $0.825\pm 0.024$ & $68.18\pm 0.38$ & $0.18\pm 0.24$ & $0.119^{+0.068}_{-0.076}$ \\

    \hdashline    
    DESI+CMB-nl+DESY3 ($3\times 2$-pt)                            
   & $0.3027\pm 0.0051$ & $0.808\pm 0.023$ & $68.28\pm 0.40$ & $0.04\pm 0.22$ & $0.044\pm 0.047$ \\
        \midrule       
   \bottomrule
    \end{tabular}
}
\caption{ 
    Constraints on modified-gravity parameters $\mu_0$ and $\Sigma_0$ from DESI (FS+BAO) data alone (with the usual BBN and $\nsten$ priors), CMB alone, and DESI in combination with external datasets. We assume the flat \lcdm\ model for the background.  We quote marginalised means and 68\% credible intervals in each case.  
    } 
    \vspace{0.1em}
    \label{tab:MG_parameter_table}
    
\end{table}
%%%%%%% END MG TABLE %%%%%%%%%%%%%%%%%%%

Whilst these MG parameterisations are a general function of $a$ and $k$, it is challenging to constrain free functions in the MG sector with current data, particularly their scale dependence. Therefore, we limit our analysis to a model with a scale-independent $\Sigma$ and $\mu$, but allow for their time (scale factor) dependence. We adopt the commonly used time dependence for the MG functions, see e.g.\ \cite{Simpson:2012ra,DES:2018ufa}, where
\begin{equation}
\mu(a)=1+\mu_{0}\frac{\Omega_{\text{DE}}(a)}{\Omega_\Lambda},\qquad
\Sigma(a)=1+\Sigma_{0}\frac{\Omega_{\text{DE}}(a)}{\Omega_\Lambda},
\label{eq:explicit-form_muSigma}
\end{equation}
where the MG parameters $\mu_0$ and $\Sigma_0$ take the value of zero in general relativity. This functional form is motivated by the desire to establish a connection between the observed cosmic acceleration and modification to gravity at late times. Consequently, the time dependence of $\mu$ and $\Sigma$ is set to be proportional to the dark energy density, Whilst these forms of time dependencies have been widely used in the literature and constitute a good basis to compare constraints across many surveys and works, they are not free from limitations and may be less effective at capturing models that depart significantly from the evolution in \cref{eq:explicit-form_muSigma}. Other parameterisations and discussions, including functional and binning forms, can be found in, e.g., \cite{Clifton:2011jh,Koyama:2015vza,Ishak:2018his}.

Finally, we note that our $\mu$--$\Sigma$ model (ansatz in \cref{eq:definition_mu} and \cref{eq:definition_Sigma}) is defined specifically in linear theory. Efforts to extend these models to non-linear scales are ongoing but are faced with multiple challenges. This is not expected to affect our results from DESI full-shape clustering and BAO, as the scale cuts in our full-shape analysis, $0.02<k/\hmpcinv<0.20$, ensure that nonlinearities (that is, the one-loop terms in the effective field theory expansion) are small. The \texttt{velocileptors} prescription used for our full-shape analysis has been tested against the MG non-linear code \texttt{fkpt} \cite{Rodriguez-Meza:2023rga}, showing a good agreement in loop corrections for small deviations from GR. The external data that we use also rely on linear scales: SN~Ia and the  primary CMB are manifestly linear; CMB lensing is almost entirely in the linear regime; and finally, for DESY3 ($3\times 2$-pt) analysis we use the same conservative scale cuts as used in the MG analysis by DES \cite{DES:2022ccp} that limits the information to linear scales. Therefore, our constraints on modified gravity rely on linear scales where this model is well-defined.

\subsection{Constraints on modified gravity}
\label{sec:MG_constraints}

Our constraints on $\mu_0$ and $\Sigma_0$ from DESI (with BBN and $\ns$ priors), and DESI in combination with external data, are presented in \cref{Fig:MG_constraints} and \cref{tab:MG_parameter_table}.  We derive our measurements assuming the \lcdm\, model for the background evolution. We find that DESI constrains $\mu_0$ to be consistent with the zero value predicted by general relativity for the motion of massive particles and their clustering, yielding:\footnote{The shaded gray region in the top left of Figure \ref{Fig:MG_constraints} results from a hard prior $\mu_0 < 2 \Sigma_0 + 1$ that is imposed when running our MCMC chains. As noted in previous works (e.g.\ \cite{DES:2018ufa,DES:2022ccp}), this is necessary to avoid an MG  parameter space where MG software based on \texttt{CAMB} encounters numerical errors when integrating the evolution of perturbations. However, this prior is of no consequence for the above results and their interpretation: while the horizontal green DESI band and the almost vertical orange CMB contour approach and just hit against this prior as shown in the figure, the other data set combinations have smaller contours and are not affected by the prior.}   
%%%%%%%%%%%%%%%
\oneonesig[6.0cm]{\mu_0 = 0.11^{+0.45}_{-0.54}}{DESI (FS+BAO)+BBN+$\nsten$}{. \label{eq:mu_DESI}} 
%%%%%%%%%%%%%%%
The marginalised mean value is centered close to the GR zero value but with $68\%$ credible interval that still allows for substantial possible deviations around it. This consistency of $\mu_0$ with zero also holds for combinations of DESI with external datasets. We see in \cref{Fig:MG_constraints} (left panel) that DESI provides no constraints on the ``lensing" MG parameter $\Sigma_0$, leading to a horizontal green  band in the $\mu_0$--$\Sigma_0$ plane. 

Adding DESI data to CMB (no lensing), and the combination of weak lensing and galaxy clustering from the Dark Energy Survey (DESY3 $3\times 2$-pt), can break degeneracies between cosmological parameters and reduce the uncertainties in $\Sigma_0$. 
The results from these combinations are consistent with the zero value predicted by GR for $\Sigma_0$ and $\mu_0$.  The combination of DESI+CMB-nl+DESY3 ($3\times 2$-pt) gives the following tight constraints on the two MG parameters:
%%%%%%%%%%%%%%%%%%%%%%%%%%%
\twoonesig[6cm]
{\mu_0    &= 0.04\pm 0.22,}
{\Sigma_0 &= 0.044\pm 0.047,} 
{DESI (FS+BAO)+CMB-nl+ DESY3 ($3\times 2$-pt). \label{eq:mu_sigma_DESI_CMB_DES3x2}}
%%%%%%%%%%%%%%%%%%%%%%%%%%%
{This result, further illustrated on the right panel of \cref{Fig:MG_constraints}, showcases the gains from adding DESI data: complementing CMB-nl and DESY3 ($3\times 2$-pt) data with DESI full-shape and BAO  improves the constraints on $\mu_0$ by a factor of 2.5,  and those on $\Sigma_0$ by a factor of 2.} 

Inspecting the CMB-only MG  constraints (see \cref{tab:MG_parameter_table}), we find the same pattern as in previous studies \cite{Planck:2015bue,Planck-2018-cosmology,DES:2022ccp,Garcia-Quintero:2020mja}, where constraints on the $\Sigma_0$ parameter from \emph{Planck} PR3 are in some tension with the zero value predicted by GR. In \cite{Planck:2015bue,Planck-2018-cosmology}, this was attributed to the anomalous amount of lensing in the CMB captured with the $A_{\rm lens}$ parameter \cite{Calabrese08,Renzi18,Mokeddem23} which we already discussed in our \cref{sec:neutrinos} above.  Adding the CMB lensing reconstruction alleviates this tension, as also found in \cite{Planck:2015bue,Planck-2018-cosmology}.   
Although this trend is not driven by DESI, the addition of DESI data to \emph{Planck} breaks parameter degeneracies and makes the  preference for nonzero $\Sigma_0$ stronger (at the 3$\sigma$ level). However, as shown by the red contours in \cref{Fig:MG_constraints} and our numbers in \cref{tab:MG_parameter_table}, we find that this tension goes away when using the more recent CMB likelihoods for low-$\ell$ \texttt{LoLLiPoP} and high-$\ell$ \texttt{HiLLiPoP} which also recently alleviated the problem of the anomalous $A_{\rm lens}$  \cite{Tristram:2021,Tristram:2023}. Therefore, our findings here shows directly that this tension is indeed linked to the lensing-anomaly issue when the \emph{Planck} PR3 likelihood is used. We illustrate and further discuss this point in our paper dedicated to detailed modified gravity analyses \cite{KP7s1-MG}.\footnote{While this paper was in DESI internal review, the paper \cite{Specogna:2024euz} appeared on the arXiv, showing similar findings regarding the $\Sigma_0$ tension being alleviated when using \texttt{LoLLiPoP} and high-$\ell$ \texttt{HiLLiPoP} likelihoods, but using a different modified-gravity software than the one we use in our analysis.}

It is worth noting that the combination of DESI with the CMB and weak lensing and galaxy clustering from the DES gives constraints on MG parameters that are comparable to those derived in \cite{DES:2022ccp} from a similar (but pre-DESI) combination of probes.\footnote{The combination in the DES extensions paper \cite{DES:2022ccp} also included SN~Ia which we do not add to our MG external-data combination, as the DESI and SN~Ia values of $\Om$ are in tension in \lcdm\ background (and remain so in the $\mu_0\Sigma_0\Lambda$CDM model).} This is due to the fact that the precision on $\mu_0$ with current DESI DR1 is approximately as powerful as that of the entire SDSS-IV dataset from two decades of observations~\cite{eBOSS:2020yzd}. It is also interesting that, while previous constraints on $\mu_0$ using SDSS-IV exhibited a tension with general relativity at a level slightly over 1$\sigma$ (see Figure 9 in \cite{DES:2022ccp}), we do not observe such a preference using DESI data.

In sum, we find that DESI full-shape data constrains the
modified-gravity parameter $\mu_0$, reflecting the sensitivity of the clustering signal to the growth of structure and the motion of massive particles where this parameter is involved. The DESI constraint on $\mu_0$ is consistent with the predictions of general relativity. Whilst DESI does not constrain the parameter $\Sigma_0$, the addition of external probes that are sensitive to gravitational lensing breaks degeneracies and produces tight constraints in the $\mu_0$--$\Sigma_0$ plane. Our constraints on modified-gravity parameters are summarised in \cref{Fig:MG_constraints} and \cref{tab:MG_parameter_table}, and a further extended analysis is presented in an accompanying paper dedicated to modified gravity \cite{KP7s1-MG}.

\section{Conclusions}
\label{sec:conclusions}

This is the second paper on cosmological results based on the 1st release (DR1) of data from the Dark Energy Spectroscopic Instrument (DESI). In the first paper \cite{DESI2024.VI.KP7A}, we presented results from the analysis of baryon acoustic oscillations in DR1. In the present paper we add information from the broadband clustering of DESI tracers, which we refer to as the ``full-shape" analysis. The major new consequence of adding the full-shape information to the baryon acoustic oscillations (BAO) is that DESI DR1 data now become directly sensitive to the temporal growth of structure, and hence to the amplitude of mass fluctuations $\sigma_8$ and other parameters that characterise cosmic growth.

Our data includes clustering from luminous red galaxies, emission line galaxies, quasars, and the \lya\ forest observed in an area of 7,500 square degrees (less for some tracers) analysed in six redshift bins in the range $0 < z < 4$. The full-shape methodology has been thoroughly validated in a series of supporting papers, leading to decisions on scale cuts and the treatment and parameterisation of systematic errors. These decisions are incorporated in the full-shape likelihood, which is subsequently combined with the BAO likelihood from \cite{DESI2024.VI.KP7A}. We refer to the results from the total likelihood as DESI (FS+BAO). When not combining with the CMB data, we complement this combination of full shape and BAO from DESI with the baryon density ($\Ob h^2$) prior from big bang nucleosynthesis, as well as a loose prior on the scalar spectral index $\ns$.

Assuming the \lcdm\ cosmological model, we find that the combination of DESI full shape and BAO pins down matter density to $\Om = 0.2962\pm 0.0095$, a $\sim$3\% measurement that is in general agreement with measurements from other cosmological probes. The amplitude of mass fluctuations is $\sigma_8=0.842\pm 0.034$, and we also constrain the derived parameter $S_8\equiv (\Om/0.3)^{0.5}=0.836\pm 0.035$; these measurements are in excellent agreement with previous galaxy-clustering analyses  as well as those from the CMB, and slightly higher than, albeit generally consistent with, constraints from weak gravitational lensing. The Hubble constant we get from DESI combined with the BBN and weak $\ns$ prior is $H_0=(68.56\pm 0.75)\kmsMpc$, in concordance with CMB and previous BAO measurements, and in continuing disagreement with the much higher values obtained by inferences from the local universe \cite{Riess:2021jrx}. 

When DESI is combined with external data while still assuming \lcdm, the constraints tighten while generally remaining in concordance with DESI-only results. The addition of the combination of cosmic shear, galaxy clustering,  and CMB lensing from the Dark Energy Survey Year-3 data --- the DESY3 ($6\times 2$-pt) analysis --- shifts the $\sigma_8$ (and $S_8$) values downward by about one standard deviation, while tightening the error bars by about a factor of 2 (and 3). When, instead, the CMB data are added to DESI, the errors in key cosmological parameters tighten even more. Finally, the most comprehensive combination that we consider in \lcdm, when DESI full-shape and BAO data are combined with the CMB and the DESY3 ($6\times 2$-pt) analysis, leads to a significant tightening of the errors and to parameter determinations with 1\% precision in $\Om$, 0.6\% in $\sigma_8$ and $S_8$, and 0.4\% precision in $H_0$.

We next study \wowacdm, the model that allows a time-varying equation of state of dark energy. The combination of full-shape and BAO data with the CMB and type Ia supernovae leads to a tightening of $\sim$20\% in the area in the $w_0$--$w_a$ plane relative to the same combinations when the full-shape information is not included. The preference for a departure from the \lcdm\ prediction ($w_0=-1, w_a=0$) remains, and we find that the best-fit $w_0$--$w_a$ model is favored by 
$\Delta \chi_\mathrm{MAP}^2=-8.8$ when DESI (FS+BAO)+CMB is combined with PantheonPlus supernovae, by $\Delta \chi_\mathrm{MAP}^2=-14.5$ for the combination with Union3, and by $\Delta \chi_\mathrm{MAP}^2=-17.5$ for the combination with DES-SN5YR (here MAP refers to the maximum \textit{a posteriori} parameter values at which the respective fits are evaluated). These correspond to preferences for  \wowacdm\ over \lcdm\ at the significance levels of $2.5\sigma$ (PantheonPlus), $3.4\sigma$ (Union3), and $3.8\sigma$ (DES-SN5YR). These combined constraints therefore continue to show preference for a departure from the \lcdm\ model at very similar statistical levels as the same combinations with DESI BAO alone in \cite{DESI2024.VI.KP7A}. We also check that the aforementioned combined-probe measurements of $\Om$, $\sigma_8$/$S_8$, and $H_0$ in the \wowacdm\ model remain consistent with those in the \lcdm\ model, with only a modest degradation in precision when the more general \wowacdm\, background is allowed.

Full-shape information allows us to improve the constraints on the sum of the neutrino masses, as neutrinos affect not only the geometry but also the growth of cosmic structure. We find an upper limit of $\sumnu < 0.409\eV$ at 95\% confidence from DESI full-shape and BAO data combined with the BBN and loose $\ns$ priors. When DESI is combined with the CMB, we obtain $\sumnu < 0.071\eV$ (again at 95\%), a constraint that is $\sim$15\% stronger than that with DESI BAO data alone combined with the CMB. This strong limit arises from the preference of both DESI BAO and DESI full-shape data for high values of $H_0$ and low values of $\Om$, which suppress the value of $\sumnu$ due to parameter degeneracy. The upper limit on $\sumnu$ is negatively correlated with the amount of lensing observed in the CMB data, and weakens if we adopt CMB likelihoods that show less evidence for  excess lensing in the CMB.

Finally, DESI full-shape data and its sensitivity to the growth of structure allow us to test the theory of gravity. We study a model where departures from general relativity are modeled by two modified-gravity parameters, $\mu_0$ and $\Sigma_0$. DESI full-shape and BAO data constrain the parameter that governs the clustering of massive particles, $\mu_0=0.11^{+0.45}_{-0.54}$, which is consistent with the zero value predicted by general relativity.  DESI alone is insensitive to the other modified-gravity parameter, $\Sigma_0$, that governs the motion of massless particles, but helps constrain it when combined with external data. DESI, in combination with CMB data, the combined lensing and clustering ($3\times 2$-pt) analysis from the DES Year-3 observations,  and DES-SN5YR supernova data, finds $\mu_0 = 0.04\pm 0.22$ and $\Sigma_0 = 0.044\pm 0.047$. These results are consistent with the zero-value predictions of general relativity. Interestingly, we find that the combined constraint on the parameter $\Sigma_0$ is a factor $4.7$ better than that on the parameter $\mu_0$ to which the full-shape analysis is sensitive,  which indicates that forthcoming DESI data should be very effective in reducing the uncertainty on the latter parameter.  

Whilst this paper wraps up the key cosmological results from DESI first data release (DR1), many ongoing or recently completed DESI projects complete the picture by studying some of the aforementioned results in more detail, or presenting complementary cosmological and astrophysical analyses. Looking ahead, BAO and full-shape analyses to follow from the three years of DESI observations are expected to contribute major new information, provide improved constraints on the cosmological parameters and models discussed in this paper, and shed new insights into dark energy, modified gravity, and neutrino mass.

\section{Data Availability}

Data from the plots in this paper will be available on Zenodo as part of DESI's Data Management Plan.

%%%%%%%%%%%%%%%%% ACKNOWLEDGEMENTS %%%%%%%%%%%%%%%%%%%%%
\acknowledgments

This material is based upon work supported by the U.S.\ Department of Energy (DOE), Office of Science, Office of High-Energy Physics, under Contract No.\ DE–AC02–05CH11231, and by the National Energy Research Scientific Computing Center, a DOE Office of Science User Facility under the same contract. Additional support for DESI was provided by the U.S. National Science Foundation (NSF), Division of Astronomical Sciences under Contract No.\ AST-0950945 to the NSF National Optical-Infrared Astronomy Research Laboratory; the Science and Technology Facilities Council of the United Kingdom; the Gordon and Betty Moore Foundation; the Heising-Simons Foundation; the French Alternative Energies and Atomic Energy Commission (CEA); the National Council of Humanities, Science and Technology of Mexico (CONAHCYT); the Ministry of Science and Innovation of Spain (MICINN), and by the DESI Member Institutions: \url{https://www.desi. lbl.gov/collaborating-institutions}. 

The DESI Legacy Imaging Surveys consist of three individual and complementary projects: the Dark Energy Camera Legacy Survey (DECaLS), the Beijing-Arizona Sky Survey (BASS), and the Mayall z-band Legacy Survey (MzLS). DECaLS, BASS and MzLS together include data obtained, respectively, at the Blanco telescope, Cerro Tololo Inter-American Observatory, NSF NOIRLab; the Bok telescope, Steward Observatory, University of Arizona; and the Mayall telescope, Kitt Peak National Observatory, NOIRLab. NOIRLab is operated by the Association of Universities for Research in Astronomy (AURA) under a cooperative agreement with the National Science Foundation. Pipeline processing and analyses of the data were supported by NOIRLab and the Lawrence Berkeley National Laboratory. Legacy Surveys also uses data products from the Near-Earth Object Wide-field Infrared Survey Explorer (NEOWISE), a project of the Jet Propulsion Laboratory/California Institute of Technology, funded by the National Aeronautics and Space Administration. Legacy Surveys was supported by: the Director, Office of Science, Office of High Energy Physics of the U.S. Department of Energy; the National Energy Research Scientific Computing Center, a DOE Office of Science User Facility; the U.S. National Science Foundation, Division of Astronomical Sciences; the National Astronomical Observatories of China, the Chinese Academy of Sciences and the Chinese National Natural Science Foundation. LBNL is managed by the Regents of the University of California under contract to the U.S. Department of Energy. The complete acknowledgments can be found at \url{https://www.legacysurvey.org/}.

Any opinions, findings, and conclusions or recommendations expressed in this material are those of the author(s) and do not necessarily reflect the views of the U.S.\ National Science Foundation, the U.S.\ Department of Energy, or any of the listed funding agencies.

The authors are honored to be permitted to conduct scientific research on Iolkam Du’ag (Kitt Peak), a mountain with particular significance to the Tohono O’odham Nation.

%%%%%%%%%%%%%%%%%%%% REFERENCES %%%%%%%%%%%%%%%%%%

\bibliographystyle{JHEP}
\bibliography{refs_key_paper,DESI2024}

%%%%%%%%%%%%%%%%% APPENDICES %%%%%%%%%%%%%%%%%%%%%

\appendix
\section{Parameter projection effects}
\label{sec:projection}

In this Appendix we give some insight into the parameter projection effects, also known as ``prior volume effects", that influence our study of the \wowacdm\ model. Projection effects typically occur in the presence of long degeneracy directions in parameter space, when the posterior exhibits strong non-Gaussianity. In those cases, the mean of the marginalised posterior can be significantly offset from the maximum of the posterior (the maximum \textit{a posteriori} (MAP) value). Strong projection effects do not indicate any problem in the analysis, but simply the fact that the statistical constraints on relevant cosmological parameters are weak due to the presence of degeneracies, which can be created by the correlation between the cosmological and nuisance parameters. We have encountered projection effects in our analysis, notably when studying the \wowacdm\ model. We have not shown in this paper any cosmological results that are subject to strong projection effects, but we now discuss one such result in order to illustrate the effect in this context. 

\begin{figure}
    \centering
    \includegraphics[width=0.95\textwidth]{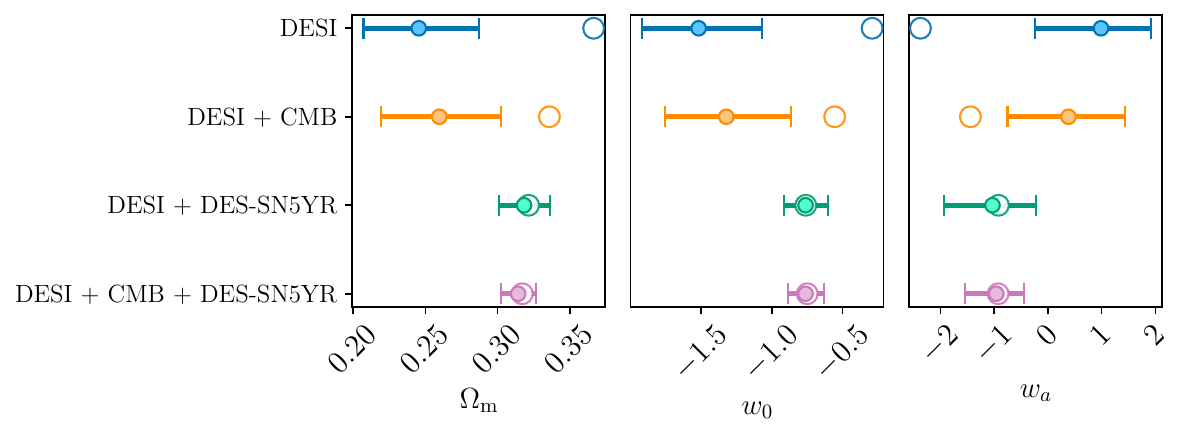}
    \caption{Parameter projection effects in the $w_0w_a$CDM cosmological model. The solid horizontal lines show the 95\% marginalised posteriors, with the solid circle being the mean of the corresponding marginalised posterior,  while the open circles show the \textit{maxima} of the corresponding posteriors (MAP values). Note that both DESI alone, and DESI in combination with CMB, have MAP values that are far from the mean of the marginalised posterior, and hence exemplify strong projection effects.  However, if DESI data are combined with type Ia supernovae (represented by the DES-SN5YR dataset here), the projection effect disappears and the marginalised means and MAP values agree well. The same result is found for the DESI+CMB+DES-SN5YR combination.  }
    \label{fig:w0wa_proj_v1}
\end{figure}

In \cref{fig:w0wa_proj_v1} we show the 1D marginalised posteriors (with 95\% credible intervals shown) on $\Om$, $w_0$, and $w_a$ in the \wowacdm\ model. We show the results for four combinations of datasets: DESI full-shape clustering and BAO data alone (with the usual BBN and $\nsten$ priors), DESI combined with the CMB, DESI combined with the DES Year-5 type Ia supernova dataset, and finally the DESI+CMB+DES-SN5YR combination. In each case and for each parameter, we also show the MAP value as an open circle. We see that the DESI and DESI+CMB combinations are strongly affected by projection effects, as the MAP values for all three parameters lie outside the corresponding Bayesian 95\% interval.
The conclusion of this analysis is that DESI (FS+BAO) --- unlike DESI BAO only ---and CMB, separately or in combination, do not give strong constraints on the dark-energy sector in \wowacdm\ model due to the presence of multiple nuisance parameters in the FS analysis.

However, once we add SN~Ia data (in this case, DES-SN5YR) to DESI, the projection effects disappear. \cref{fig:w0wa_proj_v1} shows that the MAP values are now consistent with the Bayesian posterior, due to the more effective breaking of degeneracy in the dark energy sector when SN~Ia are added to DESI, relative to the case when CMB data is added instead.  The improved constraints --- and hence lower projection effects --- with the inclusion of SN~Ia data, rather than the CMB, can be attributed to the fact the  SN~Ia are more incisive on the acceleration and complementary to galaxy clustering \cite{Linder:2005ne} and they produce tighter constraints in the $\Om$-$w_0$-$w_a$ space. Additionally, current SN~Ia observations prefer higher values of $\Om$, which further squeezes the contours in the $w_0$--$w_a$ plane. We therefore only consider the dark-energy constraints from DESI+SN~Ia (and DESI+CMB+SN~Ia etc) combinations as useful to report in \cref{sec:DE} of this paper. Further discussion of projection effects is given in the companion methodology papers \cite{DESI2024.V.KP5,KP5s2-Maus}.

%%%%%%%%%%%%%%%%% AFFILIATIONS %%%%%%%%%%%%%%%%%%%%%
% Author list file generated with: mkauthlist 1.3.0+14.gcc6daf1.dirty 
%% Affiliations file. load \usepackage{hanging}. Use \input to call it after \appendix

\section{Author Affiliations}
\label{sec:affiliations}

\noindent \hangindent=.5cm $^{1}${Instituto de F\'{\i}sica Te\'{o}rica (IFT) UAM/CSIC, Universidad Aut\'{o}noma de Madrid, Cantoblanco, E-28049, Madrid, Spain}

\noindent \hangindent=.5cm $^{2}${Lawrence Berkeley National Laboratory, 1 Cyclotron Road, Berkeley, CA 94720, USA}

\noindent \hangindent=.5cm $^{3}${Physics Dept., Boston University, 590 Commonwealth Avenue, Boston, MA 02215, USA}

\noindent \hangindent=.5cm $^{4}${Tata Institute of Fundamental Research, Homi Bhabha Road, Mumbai 400005, India}

\noindent \hangindent=.5cm $^{5}${Centre for Extragalactic Astronomy, Department of Physics, Durham University, South Road, Durham, DH1 3LE, UK}

\noindent \hangindent=.5cm $^{6}${Institute for Computational Cosmology, Department of Physics, Durham University, South Road, Durham DH1 3LE, UK}

\noindent \hangindent=.5cm $^{7}${Departamento de Astrof\'{\i}sica, Universidad de La Laguna (ULL), E-38206, La Laguna, Tenerife, Spain}

\noindent \hangindent=.5cm $^{8}${Instituto de Astrof\'{\i}sica de Canarias, C/ V\'{\i}a L\'{a}ctea, s/n, E-38205 La Laguna, Tenerife, Spain}

\noindent \hangindent=.5cm $^{9}${Department of Physics, University of Michigan, Ann Arbor, MI 48109, USA}

\noindent \hangindent=.5cm $^{10}${Leinweber Center for Theoretical Physics, University of Michigan, 450 Church Street, Ann Arbor, Michigan 48109-1040, USA}

\noindent \hangindent=.5cm $^{11}${IRFU, CEA, Universit\'{e} Paris-Saclay, F-91191 Gif-sur-Yvette, France}

\noindent \hangindent=.5cm $^{12}${Institut de F\'{i}sica d’Altes Energies (IFAE), The Barcelona Institute of Science and Technology, Campus UAB, 08193 Bellaterra Barcelona, Spain}

\noindent \hangindent=.5cm $^{13}${Instituto de Ciencias F\'{\i}sicas, Universidad Aut\'onoma de M\'exico, Cuernavaca, Morelos, 62210, (M\'exico)}

\noindent \hangindent=.5cm $^{14}${Instituto Avanzado de Cosmolog\'{\i}a A.~C., San Marcos 11 - Atenas 202. Magdalena Contreras, 10720. Ciudad de M\'{e}xico, M\'{e}xico}

\noindent \hangindent=.5cm $^{15}${Universit\'{e} Claude Bernard Lyon 1, CNRS/IN2P3, IP2I, Lyon, France}

\noindent \hangindent=.5cm $^{16}${Physics Department, Yale University, P.O. Box 208120, New Haven, CT 06511, USA}

\noindent \hangindent=.5cm $^{17}${Department of Physics and Astronomy, University of California, Irvine, 92697, USA}

\noindent \hangindent=.5cm $^{18}${Department of Physics, Kansas State University, 116 Cardwell Hall, Manhattan, KS 66506, USA}

\noindent \hangindent=.5cm $^{19}${Department of Physics \& Astronomy, University of Rochester, 206 Bausch and Lomb Hall, P.O. Box 270171, Rochester, NY 14627-0171, USA}

\noindent \hangindent=.5cm $^{20}${Institute for Astronomy, University of Edinburgh, Royal Observatory, Blackford Hill, Edinburgh EH9 3HJ, UK}

\noindent \hangindent=.5cm $^{21}${Dipartimento di Fisica ``Aldo Pontremoli'', Universit\`a degli Studi di Milano, Via Celoria 16, I-20133 Milano, Italy}

\noindent \hangindent=.5cm $^{22}${Centre for Astrophysics \& Supercomputing, Swinburne University of Technology, P.O. Box 218, Hawthorn, VIC 3122, Australia}

\noindent \hangindent=.5cm $^{23}${NSF NOIRLab, 950 N. Cherry Ave., Tucson, AZ 85719, USA}

\noindent \hangindent=.5cm $^{24}${Perimeter Institute for Theoretical Physics, 31 Caroline St. North, Waterloo, ON N2L 2Y5, Canada}

\noindent \hangindent=.5cm $^{25}${Department of Physics \& Astronomy, University College London, Gower Street, London, WC1E 6BT, UK}

\noindent \hangindent=.5cm $^{26}${Department of Astronomy and Astrophysics, University of Chicago, 5640 South Ellis Avenue, Chicago, IL 60637, USA}

\noindent \hangindent=.5cm $^{27}${Fermi National Accelerator Laboratory, PO Box 500, Batavia, IL 60510, USA}

\noindent \hangindent=.5cm $^{28}${Korea Astronomy and Space Science Institute, 776, Daedeokdae-ro, Yuseong-gu, Daejeon 34055, Republic of Korea}

\noindent \hangindent=.5cm $^{29}${Institute of Cosmology and Gravitation, University of Portsmouth, Dennis Sciama Building, Portsmouth, PO1 3FX, UK}

\noindent \hangindent=.5cm $^{30}${Department of Physics and Astronomy, University of Sussex, Brighton BN1 9QH, U.K}

\noindent \hangindent=.5cm $^{31}${Departamento de F\'{i}sica, Instituto Nacional de Investigaciones Nucleares, Carreterra M\'{e}xico-Toluca S/N, La Marquesa,  Ocoyoacac, Edo. de M\'{e}xico C.P. 52750,  M\'{e}xico}

\noindent \hangindent=.5cm $^{32}${Institute for Advanced Study, 1 Einstein Drive, Princeton, NJ 08540, USA}

\noindent \hangindent=.5cm $^{33}${Center for Cosmology and AstroParticle Physics, The Ohio State University, 191 West Woodruff Avenue, Columbus, OH 43210, USA}

\noindent \hangindent=.5cm $^{34}${NASA Einstein Fellow}

\noindent \hangindent=.5cm $^{35}${School of Mathematics and Physics, University of Queensland, 4072, Australia}

\noindent \hangindent=.5cm $^{36}${Department of Physics and Astronomy, The University of Utah, 115 South 1400 East, Salt Lake City, UT 84112, USA}

\noindent \hangindent=.5cm $^{37}${Instituto de F\'{\i}sica, Universidad Nacional Aut\'{o}noma de M\'{e}xico,  Cd. de M\'{e}xico  C.P. 04510,  M\'{e}xico}

\noindent \hangindent=.5cm $^{38}${CIEMAT, Avenida Complutense 40, E-28040 Madrid, Spain}

\noindent \hangindent=.5cm $^{39}${Department of Physics \& Astronomy and Pittsburgh Particle Physics, Astrophysics, and Cosmology Center (PITT PACC), University of Pittsburgh, 3941 O'Hara Street, Pittsburgh, PA 15260, USA}

\noindent \hangindent=.5cm $^{40}${Department of Astronomy, School of Physics and Astronomy, Shanghai Jiao Tong University, Shanghai 200240, China}

\noindent \hangindent=.5cm $^{41}${Space Sciences Laboratory, University of California, Berkeley, 7 Gauss Way, Berkeley, CA  94720, USA}

\noindent \hangindent=.5cm $^{42}${University of California, Berkeley, 110 Sproul Hall \#5800 Berkeley, CA 94720, USA}

\noindent \hangindent=.5cm $^{43}${Universities Space Research Association, NASA Ames Research Centre}

\noindent \hangindent=.5cm $^{44}${Center for Astrophysics $|$ Harvard \& Smithsonian, 60 Garden Street, Cambridge, MA 02138, USA}

\noindent \hangindent=.5cm $^{45}${Department of Physics, The Ohio State University, 191 West Woodruff Avenue, Columbus, OH 43210, USA}

\noindent \hangindent=.5cm $^{46}${The Ohio State University, Columbus, 43210 OH, USA}

\noindent \hangindent=.5cm $^{47}${Kavli Institute for Particle Astrophysics and Cosmology, Stanford University, Menlo Park, CA 94305, USA}

\noindent \hangindent=.5cm $^{48}${SLAC National Accelerator Laboratory, Menlo Park, CA 94305, USA}

\noindent \hangindent=.5cm $^{49}${Instituto de Astrof\'{i}sica de Andaluc\'{i}a (CSIC), Glorieta de la Astronom\'{i}a, s/n, E-18008 Granada, Spain}

\noindent \hangindent=.5cm $^{50}${Institute of Physics, Laboratory of Astrophysics, \'{E}cole Polytechnique F\'{e}d\'{e}rale de Lausanne (EPFL), Observatoire de Sauverny, Chemin Pegasi 51, CH-1290 Versoix, Switzerland}

\noindent \hangindent=.5cm $^{51}${Departamento de F\'isica, Universidad de los Andes, Cra. 1 No. 18A-10, Edificio Ip, CP 111711, Bogot\'a, Colombia}

\noindent \hangindent=.5cm $^{52}${Observatorio Astron\'omico, Universidad de los Andes, Cra. 1 No. 18A-10, Edificio H, CP 111711 Bogot\'a, Colombia}

\noindent \hangindent=.5cm $^{53}${Department of Physics, The University of Texas at Dallas, Richardson, TX 75080, USA}

\noindent \hangindent=.5cm $^{54}${Center for Computational Astrophysics, Flatiron Institute, 162 5\textsuperscript{th} Avenue, New York, NY 10010, USA}

\noindent \hangindent=.5cm $^{55}${Scientific Computing Core, Flatiron Institute, 162 5\textsuperscript{th} Avenue, New York, NY 10010, USA}

\noindent \hangindent=.5cm $^{56}${Institut d'Estudis Espacials de Catalunya (IEEC), 08034 Barcelona, Spain}

\noindent \hangindent=.5cm $^{57}${Institute of Space Sciences, ICE-CSIC, Campus UAB, Carrer de Can Magrans s/n, 08913 Bellaterra, Barcelona, Spain}

\noindent \hangindent=.5cm $^{58}${Departament de F\'{\i}sica Qu\`{a}ntica i Astrof\'{\i}sica, Universitat de Barcelona, Mart\'{\i} i Franqu\`{e}s 1, E08028 Barcelona, Spain}

\noindent \hangindent=.5cm $^{59}${Institut de Ci\`encies del Cosmos (ICCUB), Universitat de Barcelona (UB), c. Mart\'i i Franqu\`es, 1, 08028 Barcelona, Spain.}

\noindent \hangindent=.5cm $^{60}${Consejo Nacional de Ciencia y Tecnolog\'{\i}a, Av. Insurgentes Sur 1582. Colonia Cr\'{e}dito Constructor, Del. Benito Ju\'{a}rez C.P. 03940, M\'{e}xico D.F. M\'{e}xico}

\noindent \hangindent=.5cm $^{61}${Departamento de F\'{i}sica, Universidad de Guanajuato - DCI, C.P. 37150, Leon, Guanajuato, M\'{e}xico}

\noindent \hangindent=.5cm $^{62}${Centro de Investigaci\'{o}n Avanzada en F\'{\i}sica Fundamental (CIAFF), Facultad de Ciencias, Universidad Aut\'{o}noma de Madrid, ES-28049 Madrid, Spain}

\noindent \hangindent=.5cm $^{63}${Excellence Cluster ORIGINS, Boltzmannstrasse 2, D-85748 Garching, Germany}

\noindent \hangindent=.5cm $^{64}${University Observatory, Faculty of Physics, Ludwig-Maximilians-Universit\"{a}t, Scheinerstr. 1, 81677 M\"{u}nchen, Germany}

\noindent \hangindent=.5cm $^{65}${Department of Astrophysical Sciences, Princeton University, Princeton NJ 08544, USA}

\noindent \hangindent=.5cm $^{66}${Institut d'Astrophysique de Paris. 98 bis boulevard Arago. 75014 Paris, France}

\noindent \hangindent=.5cm $^{67}${Institute for Fundamental Physics of the Universe, via Beirut 2, 34151 Trieste, Italy}

\noindent \hangindent=.5cm $^{68}${International School for Advanced Studies, Via Bonomea 265, 34136 Trieste, Italy}

\noindent \hangindent=.5cm $^{69}${Kavli Institute for Cosmology, University of Cambridge, Madingley Road, Cambridge CB3 0HA, UK}

\noindent \hangindent=.5cm $^{70}${Department of Astronomy, The Ohio State University, 4055 McPherson Laboratory, 140 W 18th Avenue, Columbus, OH 43210, USA}

\noindent \hangindent=.5cm $^{71}${Department of Physics, Southern Methodist University, 3215 Daniel Avenue, Dallas, TX 75275, USA}

\noindent \hangindent=.5cm $^{72}${The Ohio State University, Columbus, 43210 OH, USA"}

\noindent \hangindent=.5cm $^{73}${Institute of Astronomy, University of Cambridge, Madingley Road, Cambridge CB3 0HA, UK}

\noindent \hangindent=.5cm $^{74}${Department of Physics and Astronomy, University of Waterloo, 200 University Ave W, Waterloo, ON N2L 3G1, Canada}

\noindent \hangindent=.5cm $^{75}${Waterloo Centre for Astrophysics, University of Waterloo, 200 University Ave W, Waterloo, ON N2L 3G1, Canada}

\noindent \hangindent=.5cm $^{76}${Graduate Institute of Astrophysics and Department of Physics, National Taiwan University, No. 1, Sec. 4, Roosevelt Rd., Taipei 10617, Taiwan}

\noindent \hangindent=.5cm $^{77}${Astrophysics \& Space Institute, Schmidt Sciences, New York, NY 10011, USA}

\noindent \hangindent=.5cm $^{78}${Sorbonne Universit\'{e}, CNRS/IN2P3, Laboratoire de Physique Nucl\'{e}aire et de Hautes Energies (LPNHE), FR-75005 Paris, France}

\noindent \hangindent=.5cm $^{79}${Department of Astronomy and Astrophysics, UCO/Lick Observatory, University of California, 1156 High Street, Santa Cruz, CA 95064, USA}

\noindent \hangindent=.5cm $^{80}${Department of Astronomy and Astrophysics, University of California, Santa Cruz, 1156 High Street, Santa Cruz, CA 95065, USA}

\noindent \hangindent=.5cm $^{81}${Department of Astronomy \& Astrophysics, University of Toronto, Toronto, ON M5S 3H4, Canada}

\noindent \hangindent=.5cm $^{82}${University of Science and Technology, 217 Gajeong-ro, Yuseong-gu, Daejeon 34113, Republic of Korea}

\noindent \hangindent=.5cm $^{83}${Departament de F\'{i}sica, Serra H\'{u}nter, Universitat Aut\`{o}noma de Barcelona, 08193 Bellaterra (Barcelona), Spain}

\noindent \hangindent=.5cm $^{84}${Laboratoire de Physique Subatomique et de Cosmologie, 53 Avenue des Martyrs, 38000 Grenoble, France}

\noindent \hangindent=.5cm $^{85}${Instituci\'{o} Catalana de Recerca i Estudis Avan\c{c}ats, Passeig de Llu\'{\i}s Companys, 23, 08010 Barcelona, Spain}

\noindent \hangindent=.5cm $^{86}${Max Planck Institute for Extraterrestrial Physics, Gie\ss enbachstra\ss e 1, 85748 Garching, Germany}

\noindent \hangindent=.5cm $^{87}${Department of Physics and Astronomy, Siena College, 515 Loudon Road, Loudonville, NY 12211, USA}

\noindent \hangindent=.5cm $^{88}${Department of Physics \& Astronomy, University  of Wyoming, 1000 E. University, Dept.~3905, Laramie, WY 82071, USA}

\noindent \hangindent=.5cm $^{89}${National Astronomical Observatories, Chinese Academy of Sciences, A20 Datun Rd., Chaoyang District, Beijing, 100012, P.R. China}

\noindent \hangindent=.5cm $^{90}${Steward Observatory, University of Arizona, 933 N, Cherry Ave, Tucson, AZ 85721, USA}

\noindent \hangindent=.5cm $^{91}${Aix Marseille Univ, CNRS, CNES, LAM, Marseille, France}

\noindent \hangindent=.5cm $^{92}${Departament de F\'isica, EEBE, Universitat Polit\`ecnica de Catalunya, c/Eduard Maristany 10, 08930 Barcelona, Spain}

\noindent \hangindent=.5cm $^{93}${Aix Marseille Univ, CNRS/IN2P3, CPPM, Marseille, France}

\noindent \hangindent=.5cm $^{94}${University of California Observatories, 1156 High Street, Sana Cruz, CA 95065, USA}

\noindent \hangindent=.5cm $^{95}${Department of Physics \& Astronomy, Ohio University, Athens, OH 45701, USA}

\noindent \hangindent=.5cm $^{96}${Department of Physics and Astronomy, Sejong University, Seoul, 143-747, Korea}

\noindent \hangindent=.5cm $^{97}${Abastumani Astrophysical Observatory, Tbilisi, GE-0179, Georgia}

\noindent \hangindent=.5cm $^{98}${Faculty of Natural Sciences and Medicine, Ilia State University, 0194 Tbilisi, Georgia}

\noindent \hangindent=.5cm $^{99}${Space Telescope Science Institute, 3700 San Martin Drive, Baltimore, MD 21218, USA}

\noindent \hangindent=.5cm $^{100}${Centre for Advanced Instrumentation, Department of Physics, Durham University, South Road, Durham DH1 3LE, UK}

\noindent \hangindent=.5cm $^{101}${Physics Department, Brookhaven National Laboratory, Upton, NY 11973, USA}

\noindent \hangindent=.5cm $^{102}${Beihang University, Beijing 100191, China}

\noindent \hangindent=.5cm $^{103}${Department of Astronomy, Tsinghua University, 30 Shuangqing Road, Haidian District, Beijing, China, 100190}

\noindent \hangindent=.5cm $^{104}${Physics Department, Stanford University, Stanford, CA 93405, USA}

\noindent \hangindent=.5cm $^{105}${Department of Physics, University of California, Berkeley, 366 LeConte Hall MC 7300, Berkeley, CA 94720-7300, USA}

\noindent \hangindent=.5cm $^{106}${School of Astronomy and Space Science, University of Chinese Academy of Sciences, Beijing, 100049, P.R.China}

\noindent \hangindent=.5cm $^{107}${Institute of Physics, Laboratory of Astrophysics, \'{E}cole Polytechnique F\'{e}d\'{e}rale de Lausanne (EPFL), Observatoire de Sauverny, CH-1290 Versoix, Switzerland}

\end{document}